\documentclass[two column]{aastex631}

\newcommand{\numpm}[3]{#1^{+#2}_{-#3}}
\newcommand{\unit}[1]{\,\rm{#1}}

\newcommand{\rqq}{\!>\!}

\shorttitle{}
\shortauthors{Wang et al.}

\graphicspath{{./}{figures/}}

\begin{document}

\title{Searching for new globular clusters in M\,31 with Gaia EDR3} 
\author[0000-0002-3530-3277]{Yilun Wang}
\affiliation{National Astronomical Observatories, Chinese Academy of Sciences, Beijing 100101, China}
\affiliation{School of Astronomy and Space Sciences, University of Chinese Academy of Sciences, Beijing 100049, China}

\author[0000-0003-2471-2363]{Haibo Yuan}
\affiliation{Institute for Frontiers in Astronomy and Astrophysics, Beijing Normal University,  Beijing 102206, China}
\affiliation{Department of Astronomy, Beijing Normal University, Beijing 100871,China}

\author[0000-0003-2472-4903]{Bingqiu Chen}
\affiliation{South-Western Institute for Astronomy Research, Yunnan University,  Kunming 650500, China}

\author{Xinlei Chen}
\affiliation{South-Western Institute for Astronomy Research, Yunnan University, Kunming 650500, China}

\author{Hao Wu}
\affiliation{Department of Astronomy, Beijing Normal University, Beijing 100871,China}

\author[0000-0002-3651-0681]{Zexi Niu}
\affiliation{School of Astronomy and Space Sciences, University of Chinese Academy of Sciences, Beijing 100049, China}

\author{Song Huang}
\affiliation{Department of Astronomy, Tsinghua University, Beijing 100084, China}

\author{Jifeng Liu}
\affiliation{National Astronomical Observatories, Chinese Academy of Sciences, Beijing 100101, China}
\affiliation{School of Astronomy and Space Sciences, University of Chinese Academy of Sciences, Beijing 100049, China}
\affiliation{Institute for Frontiers in Astronomy and Astrophysics, Beijing Normal University,  Beijing 102206, China}

\correspondingauthor{Haibo Yuan email:yuanhb@bnu.edu.cn}

\begin{abstract}
We found 50 new globular cluster (GC) candidates around M\,31 with Gaia Early Data Release 3 (EDR3), with the help from Pan-STARRS1 DR1 magnitudes and Pan-Andromeda Archaeological Survey (PAndAS) images.
Based on the latest Revised Bologna Catalog and \textit{simbad}, we trained 2 Random Forest (RF) classifiers, the first one to distinguish extended sources from point sources and the second one to further select GCs from extended sources.
From 1.85 million sources of $16^m{<}g{<}19.5^m$ and within a large area of $\sim$392\,deg$^2$ around M\,31, 
we selected 20,658 extended sources and 1,934 initial GC candidates. After visual inspection of the PAndAS images to eliminate the contamination of non-cluster sources, particularly galaxies, we finally got 50 candidates. These candidates are divided into 3 types (\textbf{a}, \textbf{b}, \textbf{c}) according to their projected distance $D$ to the center of M\,31 and their probability to be a true GC, $P_{GC}$, which is calculated by our second RF classifier. Among these candidates, 14 are found to be associated (in projection) with the large-scale structures within the halo of M\,31. We also provided several simple parameter criteria for selecting extended sources effectively from the Gaia EDR3, which can reach a completeness of 92.1\% with a contamination fraction lower than 10\%.
\end{abstract}

\section{Introduction}\label{introduction}

\cite{Hubble_cluster_1932} first provisionally identified 140 globular clusters (GCs) in M\,31. In this pioneering work, Hubble proposed the practical visual criteria of these extra-galactic GCs. On the photographic plates, these GCs \textit{'resemble soft, hazy star images'}, which \textit{'build up with increasing exposure in a manner perceptibly different from true stellar images.'} Meanwhile, they \textit{'appear like small condensed nebulae'} in the vision of telescopes. In addition to visual criteria, Hubble also noted specific color and radial velocity characteristics that have often been used as criteria for identifying GC candidates in M\,31 since then. \cite{bologna1988} provided a collection of catalogs, including many visually verified results over several decades\citep[see, e.g.][and references therein]{vetesnik1962,Baade1964,Sharov1973,sargent1977,crampton1984,battistini1980,battistini1987}, while \cite{RBC2004} revised the Bologna Catalog \citep{battistini1987} with the Two Micron All Sky Survey (2MASS), resulting in a catalog that included 337 confirmed GCs and 688 GC candidates. This Revised Bologna Catalog (RBC) is kept updated till 2012 to the 5th edition, the RBC V\,5\footnote{For detail, please refer the RBC website,http://www.bo.astro.it/M31/}, which contains 625 GCs and 331 GC candidates. The Panchromatic Hubble Andromeda Treasury (PHAT) survey used the Hubble Space Telescope (HST) to provide high-resolution images of one-third of the M\,31 disk, leading to the discovery of 2753 stellar clusters through the Andromeda Project (AP)\citep{PHAT2015}, which involved over 1.8 million classifications made by tens of volunteers through the internet. \cite{Huxor2014} reported 59 GCs and 2 GC candidates in a search of the halo of M\,31 using images from the Pan-Andromeda Archaeological Survey (PAndAS). while \cite{WSC_2022} utilized a convolution neural network (CNN) to identify 117 new M\,31 GC candidates from 21 million PAndAS source images.

Although visual inspection is always the most important criterion for GC identification, the tedious work of candidate selection needs to be accelerated with some high-efficiency methods rather than visual inspection. Except for the machine learning methods with the images \citep{WSC_2022}, it is more common to utilize the catalog data to select the candidates, like the photometry information and the morphology parameters. For example, the non-stellar flag used in \cite{Huxor2014} and the full width at half-maximum (FWHM) of the point spread function (PSF) used in \cite{NVK2008}. 

 The Gaia satellite scans the entire sky from the second Lagrange (L$_2$) point of the Sun-Earth-Moon system and has a spatial resolution comparable to that of HST \citep{gaia_mission}. Thus the Gaia catalog contains a huge treasure of high-resolution image information that can be used to search for extended sources. For example,  \cite{voggel2020} used parameter cuts with Gaia Data Release 2 (DR2) to identify 632 new candidate luminous clusters in the halo of the elliptical galaxy Centaurus A. People also used the existed spectroscopic catalogs as labeled input data to train the machine learning model on Gaia DR2 or DR3, for source classification, like star/galaxy/qso, and etc  \citep{bailer-jones-qso-galaxy,gaia_dr3_non_stellar}.  

In this work, we utilized the catalog data from Gaia EDR3 and Pan-STARRS1 (PS1) to search for GC candidates in the large area centered around M\,31. We then conducted a visual check using images from PAndAS. The data and method used are described in Sections~\ref{section:data} and \ref{section:process}, respectively, and the results are presented in Section~\ref{section:result}. Specifically, our method involves a two-step machine learning model based on catalog data. Finally, we summarize and conclude in Section~\ref{section:summary_discussion}

\section{Data}\label{section:data}
\subsection{Pan-STARRS1}
The Panoramic Survey Telescope and Rapid Response System (Pan-STARRS), located at Hawaii, conducted the stacked $3\pi$ Steradian Survey in 5 broad band $grizy_{p1}$. The $5\sigma$ stack limits for these bands are $(23.3,23.2,23.1,22.3,21.4)$, while the bright limits are $(14.5,15.0,15.0,14.0,13.0)$\citep{ps1}. We used the PSF magnitude of $grizy_{p1}$, which was calculated within a PSF-matched radial aperture that was adjusted for each source by the pipeline software \textbf{\textit{psphot}}\citep{ps1-pixel}. Additionally, we used the colors  $(g{-}r,r{-}i,i{-}y,y{-}z)$.

\subsection{Gaia EDR3}\label{section:gaia_data}
The Gaia processing treats all sources as single stars, which underestimates the $G$-band magnitude of the extended sources(please see Figure 2 in \cite{gaia_dr3_non_stellar}). According to \cite{gaia_lsf}, a source passes through the CCDs of the Gaia satellite and receives a window determined by its brightness. All sources in the Gaia vision are treated as single stars, meaning that extended sources receive the same window as point sources. The Astrometry Field (AF) is the primary part of the CCD, which collects astrometry and $G$-band photometry information. Gaia CCDs have two directions: the along-scan (AL) and across-scan (AC) directions. As the instantaneous spatial resolution in the AL direction of Gaia is comparable to that of HST \citep{gaia_mission}, the window size in this direction is 0.7 arcsec (12 pixels) for sources with $16{<}G{<}20.7$, which increases to 1 arcsec (18 pixels) for sources with $13{<}G{<}16$. For sources with $G{>}13$, the pixels are binned in the AC direction. Within this window, data is fitted with a line spread function (LSF) to obtain the centroid position (astrometry) and flux ($G$-band photometry) \citep{astrometric_solution_dr2,gaia_lsf}. However, the fixed window and LSF fitting method may produce different results in position and $G$ magnitude for extended sources, particularly those larger than 0.7 arcsec. On the other hand, the magnitudes in $BP$ and $RP$ bands are aperture magnitude generated from a larger window with 60 pixels in the AL direction, which can obtain more light and avoid fitting the LSF. In addition to magnitude in the $G/BP/RP$ band, we also used the following parameters, 
\begin{enumerate}
    \item $bp\_rp\_excess$: phot\_bp\_rp\_excess\_factor, which measures the excess flux of $BP$ and $RP$ with respect to $G$  \citep{gaia_photometry_2021}. 
    
    \item $astrometric\_excess\_noise$: This parameter measures the difference between the best-fitting astrometric model (model for single point source) and the observations of a source. It implies the observational noise or the residuals in the fitting.\citep{gaia_doc_20,astrometric_core_solution}
    
    \item $pos\_err$: We define this parameter by $pos\_err=\sqrt{\sigma_{RA}^2+\sigma_{DEC}^2}$, which is slightly different from the $\sigma_{pos\_max}$ \citep{astrometric_solution_dr2}, for the convenience in cross-matching with \textsf{TOPCAT} \citep{topcat}.
    
    \item $astrometric\_params\_solved$: In Gaia EDR3, there are 3 kinds of astrometric solutions for 2-parameter, 5-parameter, and 6-parameter. Then the parameter $astrometric\_params\_solved$ is (3, 31, 95) respectively. More details about these three models can be found in \citep{gaiadr3_document}. The 5-parameter solution usually gets higher precision, while the 6-parameter solution has a larger uncertainty. For the 2-parameter solution, only the RA and Dec are provided, and both the proper motion and parallax are omitted. 
    
    \item parallax$(\varpi)$ and parallax\_error$(\sigma_{\varpi})$: If a source only get the 2-parameter solution, we set the $\varpi=-9999$ and the $\sigma_{\varpi}=-10$. 
\end{enumerate}
In this work, we often used the logarithm form of some parameters, including $\log(bp\_rp\_excess)$, $\log(pos\_err)$, and $\log(astrometric\_params\_solved)$.

\subsection{LAMOST DR8 LRS}\label{section:data_lamost}
We used the classification data in the Low-Resolution Spectroscopic Survey of the Large Sky Area Multi-Object Fiber Spectroscopic Telescope (LAMOST) as auxiliary data in Sections~\ref{section:regression} and \ref{section:parameter_distribution}. The LAMOST, also called Guo Shou Jing Telescope, is a specially designed Schmidt telescope with both a large aperture (effective aperture of 3.6–4.9m) and a wide field of view ($5^\circ$). It is capable of observing thousands of targets using 4,000 fibers in a single observation \citep{LAMOST_2012}. The spectrum wavelength coverage is from 370 to 900$\,\mathrm{nm}$, with a resolving power of ${\sim}1800$ \citep{LAMOST_overview}. Based on \textbf{\textit{specBS}} used by Sloan Digital Sky Survey (SDSS), the LAMOST spectrum analysis pipeline uses a method of principal component analysis to classify the spectra and measure the radial velocity \citep{LAMOST_pilot,LAMOST_DR1}. The LAMOST DR8 LRS provides spectral classifications for each source, including Star/Galaxy/QSO/Unknown. We used a sub-sample of LAMOST DR8 LRS that had a signal-to-noise ratio larger than 20. To avoid contamination from M\,31, we excluded the sources in a $3.5^{\circ}\times3.5^{\circ}$ area centered at M\,31. 

\subsection{PAndAS}
Our visual inspection relies on the optical images of the PAndAS survey, which were obtained around M\,31 and M\,33 by the Canada–France–Hawaii Telescope (CFHT), with an average seeing of 0.6 and 0.67 arcsec in $i$ and $g$ band, respectively. To investigate the stellar halo of M\,31, the PAndAS survey covers a large area (${>}400\ \rm{deg}^2$) that reaches a projected distance of $D{\sim}150\unit{kpc}$ from the center of M\,31 \citep{pandas_structure_2014,pandas_structure_2018}.

\subsection{Total sample}
We selected a area around M\,31 (R.A.=00 42 44.330, Dec=+41 16 07.50), with R.A. in  $0^\circ{\sim}22.26^\circ$ and Dec in  $29.26^\circ{\sim}53.26^\circ$. We imposed a magnitude restriction in the Pan-STARRS $g$-band. Based on the RBC V5, which shows that most GCs have magnitudes in the range of $16^{m}{<}g{<}21^{m}$, we set the bright limit at $16^{m}$. The faint limit was set at $19.5^{m}$ to enable the use of $G$-band photometry-related parameters and to reduce contamination at the faint end. We might lose some possible GC candidates in the range of $19.5^{m}{<}g{<}21^{m}$, since they might be indiscernible to our method. It was observed that 94.9\% of the 492 globular clusters (GCs) located in RBC V5 exhibit $g-r>0$. When we take into account the GCs that are located outside the disk of M\,31, specifically those at a distance $D>0.9^\circ$ from the center of M\,31, all of the GCs show $g-r>0$. Therefore, we imposed a color restriction of $g{-}r{>}0$. These selection criteria resulted in a total sample of 1.85 million sources obtained from cross-matching between PS1 and Gaia EDR3 using a matching radius of $1''$.

\subsection{Training sample}\label{section:training_sample}
In Sections~\ref{section:classify1} and \ref{section:classify2}, we used a training sample based on RBC V5, which included sources labeled as globular clusters, galaxies, and stars. The sources with no PS1 data or Gaia EDR3 data were discarded. In addition, we also cross-matched our total sample with \textit{simbad} and added the sources with \textit{simbad} parameter $main\_type$ being [Star, EB*, Galaxy, GlCl]. Here, EB* refers to eclipsing binary, also labeled $star$, and GlCl is globular cluster. If a source was labeled differently in RBC V.5 and \textit{simbad}, we retained the \textit{simbad} $main\_type$. Our resulting training sample consisted of 12,343 sources, including 8,024 stars, 3,876 galaxies, and 443 GCs.


\section{Methods}\label{section:process}
\begin{figure}[h]
\centering
\includegraphics[scale=0.124]{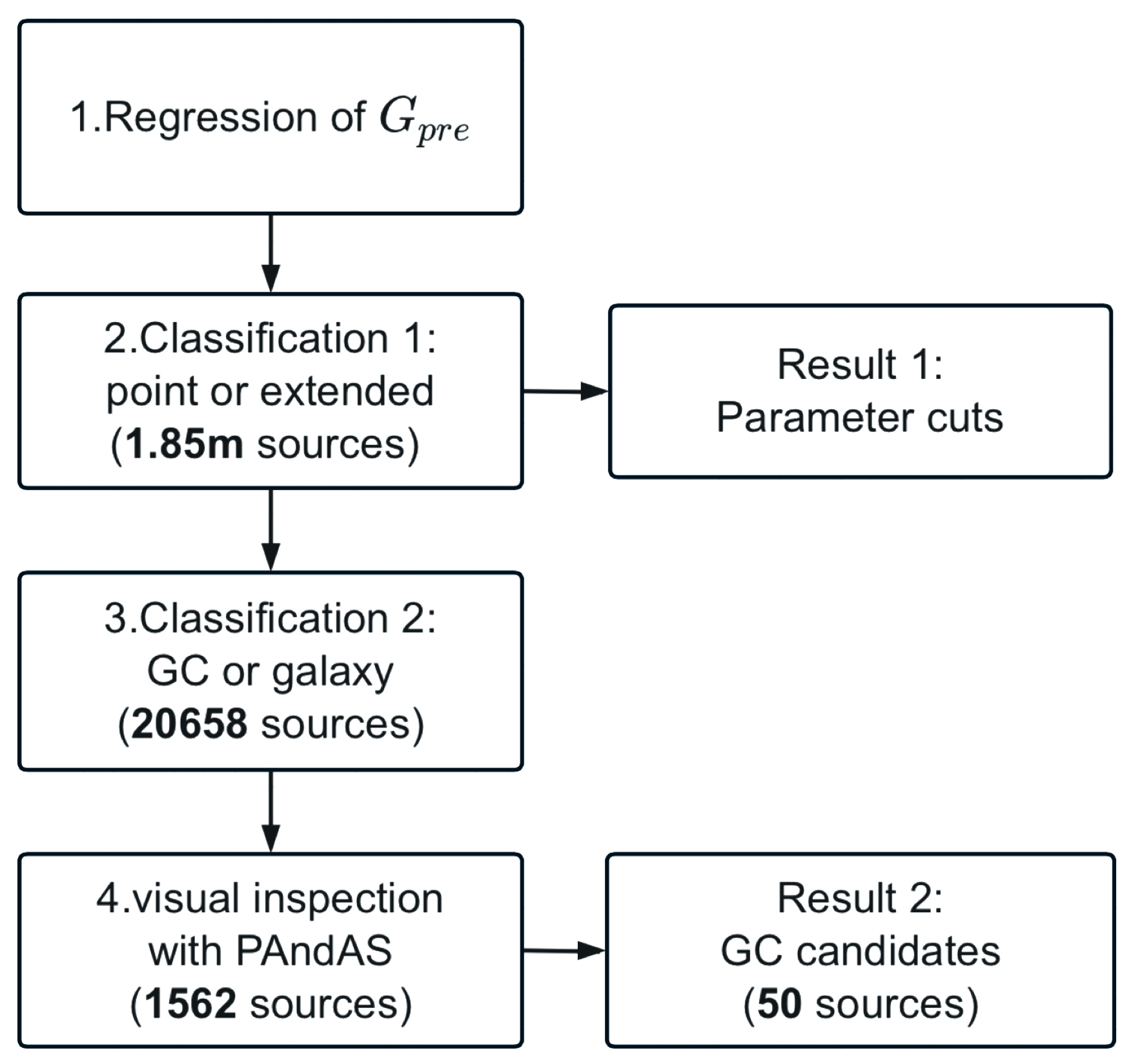}
\caption{\label{fig:process_method}The flow chart of Section~\ref{section:process}.}
\end{figure}

The main process of searching for GC candidates is illustrated in Figure~\ref{fig:process_method}, which consists of 4 steps and 2 main results. In Step 1, we performed a regression of the $G$ magnitude on the Pan-STARRS colors $(g{-}r,r{-}i,i{-}y,y{-}z)$ and $r$ magnitude in Section~\ref{section:regression}, resulting in a new parameter for use in subsequent classification. Step 2 employed a random forest method to distinguish point and extended sources in Section~\ref{section:classify1}. In Step 3, we tried to remove the galaxy contamination in the extended sources ($P_{ext}{>}0.5$) with another random forest classifier in Section~\ref{section:classify2}. Finally, in Section~\ref{section:visual_inspec}, we discuss the Step 4 of visual inspection with PAndAS images.

\subsection{Regression of $G$ with PS1 photometry}\label{section:regression}
We used the subsample in Section~\ref{section:data_lamost} as point sources, which consists of approximately 340,000 sources that meet the same magnitude and color criteria as the total sample. Table~\ref{tab:lamost_color} displays the differences in the four colors $(g{-}r,r{-}i,i{-}z,z{-}y)$ between stars and galaxies, which are typically about 0.1–0.2. When comparing the Gaia $G$ magnitude and the Pan-STARRS $r$ magnitude, we found that the color $G{-}r$ for stars is 0.01, similar to the $G{-}r$ value of 0.03 for QSO, while it reaches 2.26 for galaxies. This large difference between point sources (stars) and extended sources (galaxies) is caused by the PSF magnitude in $G$-band of Gaia (see Section \ref{section:gaia_data}), which is different from the aperture magnitude in $grizy_{p1}$-band of Pan-STARRS and $BP/RP$-band of Gaia. The application of PSF fitting method to the extended sources within a small Gaia window might cause some problem in deriving the magnitude. Hence, it appears that $G$ is accurate for point sources but may be incorrect for extended sources. Therefore, We anticipate that this regression could generate a better magnitude in $G$-band for extended sources which might be used in our classifications. In addition, since $G{-}r$ value for point sources is mostly around 0, we added $G{-}r<1$ as an additional criterion for the regression sample, which represents a pure point source sample.  

\begin{deluxetable}{cccc}[ht]
\tablecaption{The median of different colors of the 3 types of sources in the subsample of LAMOST DR8 LRS.\label{tab:lamost_color}}
\tablewidth{0pt}
\tablehead{
\colhead{Color} &\colhead{STAR}&\colhead{GALAXY}
&\colhead{QSO}\\
\colhead{ } &\colhead{(320,000)}&\colhead{(18,618)}
&\colhead{(963)}}
\startdata
$g{-}r$ & $\numpm{0.57}{0.50}{0.22}$ & $\numpm{0.44}{0.49}{0.24}$ & $\numpm{0.21}{0.15}{0.14}$ \\
$r{-}i$ & $\numpm{0.23}{0.27}{0.10}$ & $\numpm{0.34}{0.13}{0.16}$ & $\numpm{0.11}{0.24}{0.15}$ \\
$i{-}z$ & $\numpm{0.09}{0.14}{0.06}$ & $\numpm{0.24}{0.13}{0.14}$ & $\numpm{0.02}{0.20}{0.13}$ \\
$z{-}y$ & $\numpm{0.07}{0.08}{0.04}$ & $\numpm{0.21}{0.12}{0.12}$ & $\numpm{0.06}{0.14}{0.18}$ \\
$G{-}r$ & $\numpm{0.01}{0.04}{0.02}$ & $\numpm{2.26}{0.34}{0.36}$ & $\numpm{-0.03}{0.24}{0.14}$
\enddata 
\tablecomments{The number in the bracket under the column head is the number of sources used in this table. The error in this table is the $16^{\rm{th}}$ and $84^{\rm{th}}$ percentile.}
\end{deluxetable}

We used the stochastic gradient descent method \textsf{SGDregressor} of scikit-learn \citep{scikit-learn} to do the regression, which is well suited for a large training sample. We regressed the $G{-}r$ on the 4 Pan-STARRS color, resulting in a predicted color, denoted $(G{-}r)_{SGD}$. Using this predicted color, we calculated the predicted $G$ magnitude via $G_{pre}{=}r{+}(G{-}r)_{SGD}$.We then used the magnitude difference, $G_{pre}-G$, in subsequent steps of our analysis.

\subsection{Classification 1: point or extended}\label{section:classify1}
In our first classification, we tried to distinguish the point source and the extended source. We used the Random Forest (RF) method with 200 estimators. To select the most relevant parameters, we ran the RF with the training sample 200 times and collected the mean parameter importance $imp_1$ of the 21 parameters in Table.\ref{tab:parameter importance}. We chose the parameters with $imp_1{>}0.03$ to generate the point-extended classifier, Clf1. This classifier identified 20,658 extended sources out of the 1.85 million sources in the total sample and assigned each source a probability of being an extended source, $P_{ext}$. To assist in distinguishing extended sources from point sources, we presented a set of parameter cuts based on the results from Clf1 in Table \ref{tab:cuts_predicted}. Additionally, for comparison, we provided another set of parameter cuts in Table \ref{tab:cuts_lamost}, utilizing the classification information from LAMOST LRS DR8. Further details and discussions are presented in Section \ref{section:parameter_distribution}.

\begin{deluxetable}{cccc}[ht]
\tablecaption{The importance of the parameters used in the Cl1 and Clf2 in Section~\ref{section:classify1} and  Section~\ref{section:classify2}.\label{tab:parameter importance}}
\tablewidth{0pt}
\tablehead{
\colhead{Parameter} &\colhead{$imp_1$}&\colhead{$imp_2$}
&\colhead{Used}}
\startdata
$\log\,(bp\_rp\_excess)$ & 0.200 & 0.044 & 1,2 \\
$G_{pre}{-}G$ & 0.182 & 0.045 & 1,2 \\
$G{-}r$ & 0.150 & 0.041 & 1,2 \\
$\log\,(pos\_err)$ & 0.114 & 0.054 & 1,2 \\
$\log\,(astrometric\_excess\_noise)$ & 0.088 & 0.068 & 1,2 \\
$G$ & 0.073 & 0.059 & 1,2 \\
parallax ($\varpi$) & 0.054 & 0.011 & 1 \\
parallax\_error ($\sigma_{\varpi}$) & 0.038 & 0.007 & 1 \\
$astrometric\_params\_solved$ & 0.035 & 0.003 & $-$ \\
$BP{-}g$ & 0.013 & 0.034 & 2 \\
$i{-}z$ & 0.009 & 0.095 & 2 \\
$g$ & 0.007 & 0.050 & 2 \\
$r$ & 0.006 & 0.035 & 2 \\
$i$ & 0.005 & 0.044 & 2 \\
$z$ & 0.004 & 0.066 & 2 \\
$BP$ & 0.004 & 0.057 & 2 \\
$y$ & 0.004 & 0.066 & 2 \\
$RP$ & 0.004 & 0.037 & 2 \\
$g{-}r$ & 0.003 & 0.043 & 2 \\
$r{-}i$ & 0.003 & 0.039 & 2 \\
$z{-}y$ & 0.003 & 0.101 & 2 
\enddata
\tablecomments{Col1 is the name of the parameters. Col2 is the parameter importance of Clf1 while Col3 is the parameter importance of Clf2. Col4 shows which classifier used the corresponding parameter. In Col4, '1,2' means that a parameter was used in both classifiers. while '-' means that we didn't use it in our classifiers. }
\end{deluxetable}

\subsection{Classification 2: cluster or galaxy}\label{section:classify2}
To distinguish between the GCs and galaxies,  we used the same training procedure as described in Section \ref{section:classify1} on the subsample of GC/G in the training sample. We also calculated the parameter importance $imp_2$ and selected 17 parameters with $imp_2 \rqq 0.03$ from Table~\ref{tab:parameter importance} to generate the GC/G classifier, Clf2. The Clf2 gives the probability of being a GC, $P_{GC}$, for all the 20,658 extended sources, from which we selected 1,934 sources with $P_{GC}{>}0.25$ to cross-match with the PAndAS image. We then obtained images of 1,562 sources to perform the visual inspection, while other sources got no PAndAS image.

\subsection{visual inspection}\label{section:visual_inspec}

\begin{figure}[h!]
\centering
\includegraphics[scale=0.72]{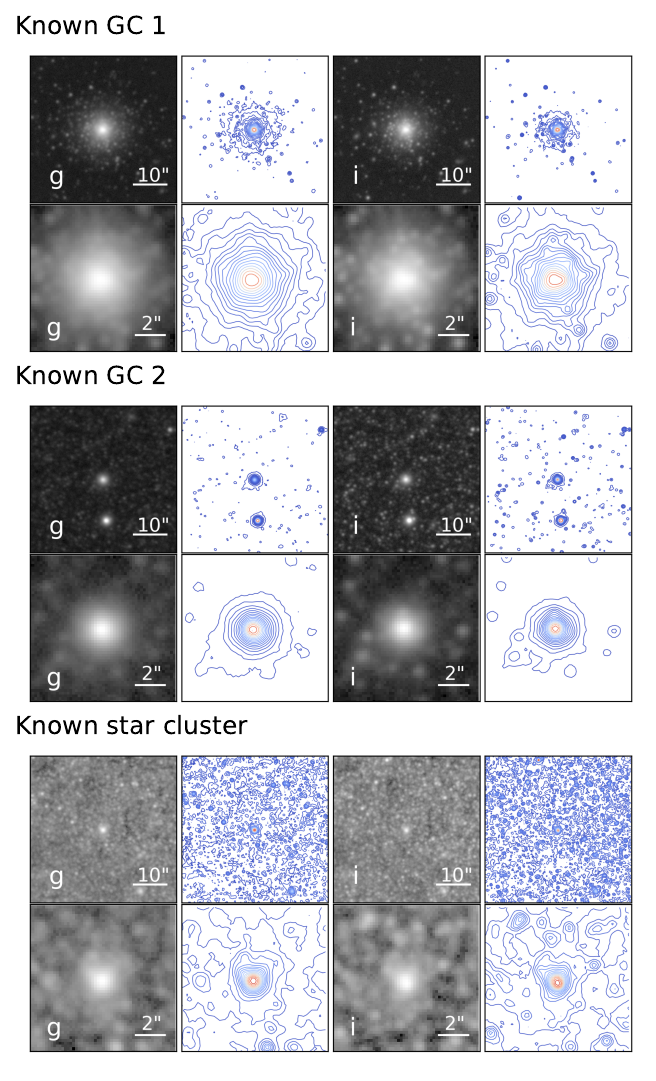}
\caption{\label{fig:known_GCs}We plot the images and contours of 3 typical clusters in 3 panels. In the first panel, the $240{\times}240$ pixels images of a typical GC in $g$ and $i$ band are displayed in the first row, while the zoom-in $54{\times}54$ pixels images are shown in the second row. Adjacent to these images, the contour plots are shown in the same scale. We exhibit another known GC with a comparatively smaller size in the second panel. A known star cluster from \cite{PHAT1} are shown in the third panel.}
\end{figure}

\begin{figure}[h!]
\centering
\includegraphics[scale=0.56]{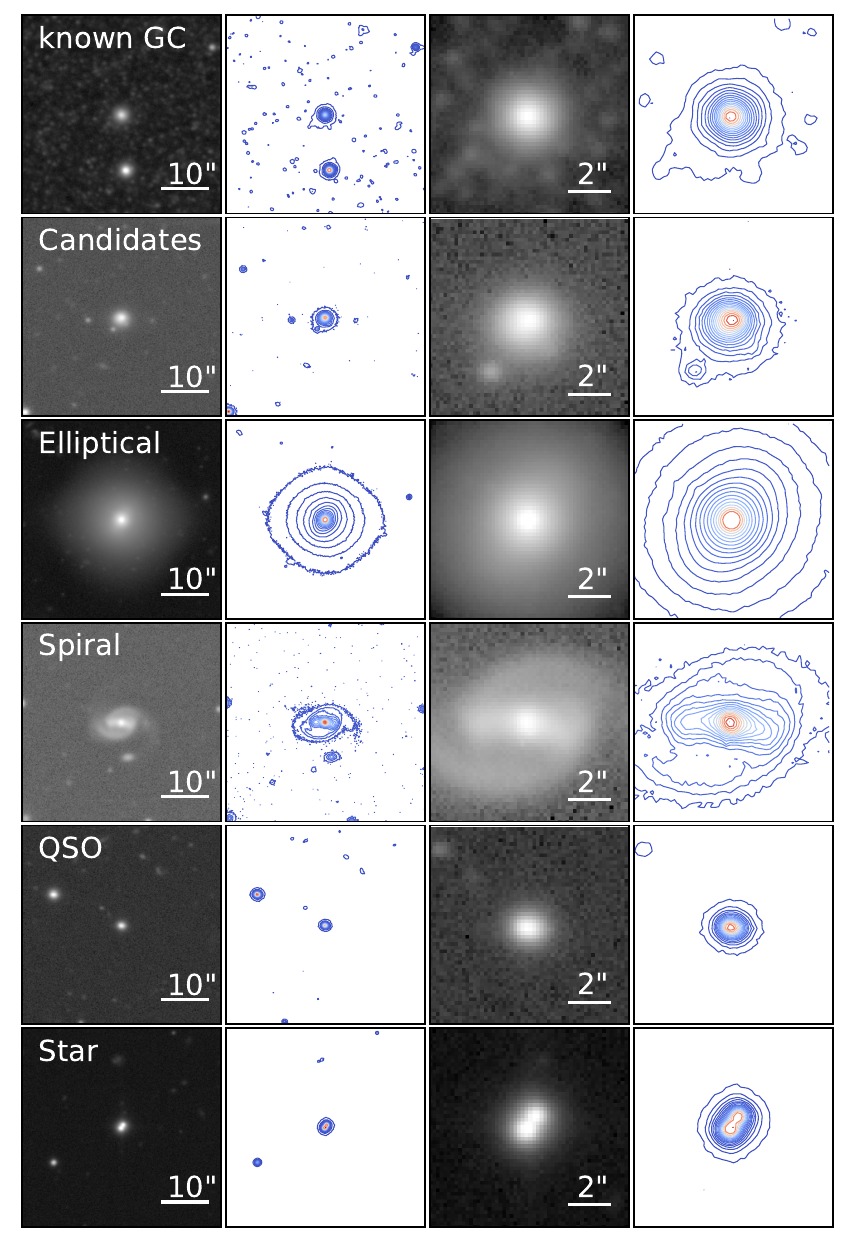}
\caption{\label{fig:cm_fig_type} PAndAS $g$-band images and contours of 6 typical types of sources in the extended sources with $P_{GC}{>}0.25$, including the known GC, GC candidates, elliptical galaxy, spiral galaxy, QSO and star. The 1st column shows the $240{\times}240$ pixels images, while the 3rd column shows the zoom-in $54{\times}54$ pixels images. The 2nd and 4th columns display the contour of column 1 and column 3, respectively.} 
\end{figure}

In order to identify potential globular cluster (GC) candidates and mitigate contamination, we generated PAndAS images and contours for the remaining 1562 extended sources, as depicted in Figure \ref{fig:known_GCs}. These images had dimensions of $240{\times}240$ pixels ($44.87{\times}44.87\ \rm{arcsec^2}$) and $54{\times}54$ pixels ($10.1{\times}10.1\ \rm{arcsec^2}$) in the $g$ and $i$ bands, respectively. Each image was examined carefully by an experienced astronomer in conjunction with relevant catalog information, including the \textit{simbad} classification, $P_{ext}$, and $P_{GC}$. The results were checked by two astronomers, and a consensus was reached. Out of the 1562 sources, 592 possessed a \textit{simbad} classification, including 388 known GCs. Additionally, we plotted a subset of sources with $P_{GC}{<}0.25$ to familiarize the inspector with the characteristics of these extended sources, as most galaxies were excluded using the criterion $P_{GC}{>}0.25$. For instance, among the 4670 extended sources with a \textit{simbad} main\_type indicative of galaxies, a noteworthy 97.5\% exhibited $P_{GC}{<}0.25$. Based on the image features and other pertinent information, we mitigated most of contamination and identified the probable GC candidates.

In Figure\,\ref{fig:known_GCs}, we present the morphologies of three distinct types of known clusters across three panels. The first panel features a example of M\,31 GCs (Gaia DR3 361274923609629696), which exhibits a larger diameter compared to the other two types of clusters. It displays numerous distinct luminous points arranged in an irregular contour within a bright circular area.
In the second panel, we showcase another known GC (Gaia DR3 381337059455029888) that appears more compact than the previous one. It manifests as a bright circle with few or no bright sources found within it. Despite its more condensed nature, this GC still exhibits an extended structure in the $240{\times}240$ pixel image, surpassing the size of other point sources within the same field of view.
The third panel depicts a known star cluster (Gaia DR3 387310328163704960) sourced from \cite{PHAT1}. This star cluster is represented as a circular feature with a diameter slightly larger than $2''$, just marginally exceeding the dimensions of point sources in the $54{\times}54$ pixel image.

In Figure\,\ref{fig:cm_fig_type}, we illustrate the primary types of contamination, which include elliptical galaxies, spiral galaxies, QSOs, and stars. To provide a comprehensive comparison, we also plot a known globular cluster (GC) and a GC candidate within the same figure. We briefly describe these sources of contamination as follows: a) Elliptical galaxies typically exhibit a prominent central core with a large extended profile. Their contour plots often depict regular but non-circular shapes. b) Spiral galaxies tend to resemble ellipses and contain distinctive features such as bars and arms. c) QSOs are compact extended sources, smaller in size compared to other contamination types. d) Stars, which are primarily binary stars in our contamination scenario, can be resolved by the PAndAS images.

During our analysis, we found that all previously known globular clusters (GCs) with the similar image of the known GC 1 in Figure\,\ref{fig:known_GCs}, had already been identified in prior studies. Consequently, our focus shifted towards identifying candidates exhibiting similar features in both image and contour to GC 2 in Figure\,\ref{fig:known_GCs}. We found it straightforward to recognize the characteristic traits of spiral galaxies and stars. For elliptical galaxies, we relied on the distinguishing attributes of a large profile and non-circular contours, which differentiate them from known GCs. To further differentiate between several QSOs and GC candidates, we employed $P_{ext}$, where the QSOs exhibited $P_{ext}=0.608^{+0.011}{-0.079}$ compared to the GC candidates with $P{ext}=0.995^{+0.005}_{-0.071}$.

\begin{figure*}
\centering
\includegraphics[scale=0.21]{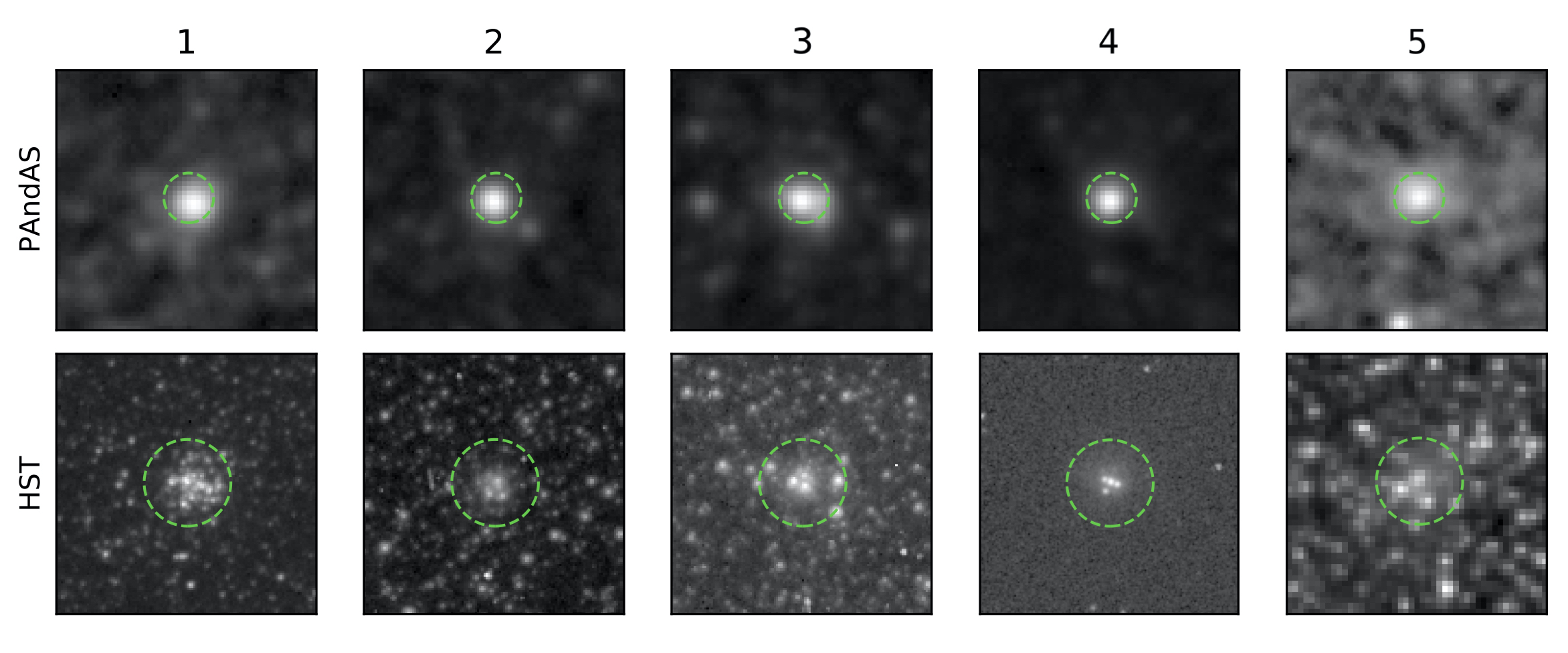}
\caption{\label{fig:pandas_hst}The PAndAS $g$-band images (top panels) and HST (bottom panels) images of 5 possible cluster candidates in our 54 visual checked candidates. The HST images were taken using the following filters: F475W, F814W, F814W, F275W, and F127M, respectively. Upon examining the HST images, we eliminated sources 2-5 from our final catalog of candidates. In both sets of panels, the green dashed circles indicate a radius of $1''$, with the center aligned to the J2000 position of each respective source.}
\end{figure*}

After the inspection of PAndAS images, we got 54 visual checked candidates. Among these, we obtained HST images for five candidates located within the M\,31 disk. In Figure\,\ref{fig:pandas_hst}, source 1 exhibits an appearance of a stellar cluster, while sources 2-5 contain fewer stars within a $1''$ radius compared to source 1. Considering that source 1 is identified as a GC candidate in the RBC V5 catalog and as a stellar cluster candidate in \cite{NVK2008}, we decided to retain it in our candidate catalog. However, sources 3-5 were excluded from the catalog as they were classified as star clusters or HII regions in previous studies. Source 2, in particular, displayed a similar image to source 3, and thus it was also excluded for consistency. For the remaining candidates outside the disk, no corresponding HST images were available, limiting our ability to verify them at a higher resolution.

\begin{deluxetable*}{cccccccc}[ht!]
\tablecaption{The information of the HST images in Figure\,\ref{fig:pandas_hst}}\label{tab:hst_candidates}
\tablewidth{0pt}
\tablehead{
\colhead{Source ID\tablenotemark{a}} & \colhead{RA} & \colhead{Dec} & \colhead{Previous Candidate\tablenotemark{b}} & \colhead{Instrument\tablenotemark{c}} & \colhead{Filter} & \colhead{Dataset}& \colhead{Observation Date}
}
\startdata
 1& 00:41:27.01 & +40:41:37.26 & Yes\tablenotemark{d} & ACS & F475W &  JEPU06020 & 2021-12-22\\
 2& 00:42:32.68 & +40:57:38.17 & No & ACS & F814W & J92GC4Q0Q & 2005-02-18\\
 3& 00:40:15.37 & +40:36:54.64 & Yes\tablenotemark{e} & ACS & F814W & JEPV30010 & 2022-07-15\\
 4& 00:41:02.00 & +41:02:54.79 & Yes\tablenotemark{f} & ACS & F275W & ID5805010 & 2017-06-25\\
 5 & 00:42:33.04 & +41:10:52.05 & Yes\tablenotemark{g} & WFC3 & F127M & IE0X74010 & 2020-12-06
\enddata
\tablenotetext{a}{The Source ID is the same ID in Figure~\ref{fig:pandas_hst}.}
\tablenotetext{b}{Whether a source was labeled as a GC candidate in previous research.}
\tablenotetext{c}{The Instrument used on HST to acquire data.}
\tablenotetext{d}{It was classified as a GC candidate in RBC V5, and as a stellar cluster candidate in \cite{NVK2008}. }
\tablenotetext{e}{It was classified as a stellar cluster in \cite{NVK2008}.}
\tablenotetext{f}{It was classified as an HII region \cite{HII_2011}.}
\tablenotetext{g}{It was classified as a GC candidate in \cite{KPNO_2010}, and as a star cluster in \cite{HST_cluster_WFPC2}.}
\end{deluxetable*} 
\section{Result and Discussion}\label{section:result}
First, we present the result of the regression and the classification in Section~\ref{section:result_regression_classification}. Then in Section~\ref{section:parameter_distribution}, we show the parameter distribution for the point and extended sources, and provide a series of parameter cuts to select the extended sources. In Section~\ref{section:result_point_extend}, we discuss the parameter cuts and a small part of sources confused by Clf1 . In Section~\ref{section:result_compare_catalogs}, the extended sources predicted by Clf1 are compared with some existing catalogs. The 50 GC candidates are discussed in Section~\ref{section:result_candidates}.
\subsection{Result of regression and classification}\label{section:result_regression_classification}
\begin{figure}[h]
\centering
\includegraphics[scale=0.57]{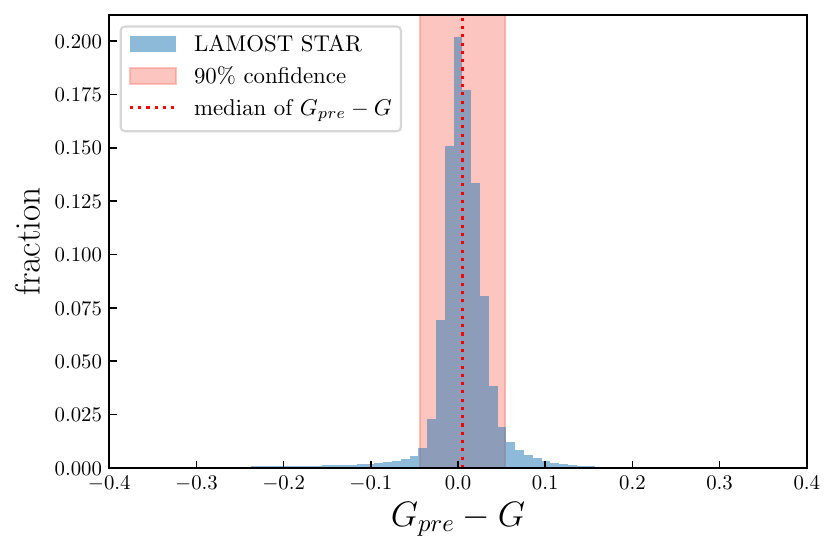}
\caption{ \label{fig:gdiff_dis}The distribution of the magnitude difference of the LAMOST stars between the predicted $G$ band magnitude, $G_{pre}$, and the Gaia $G$ magnitude. The red shadow shows a range of 90\% confidence. The red dashed line indicates the median of $G_{pre}{-}G$.}
\end{figure}

\begin{figure}[h!]
\centering
\includegraphics[scale=0.66]{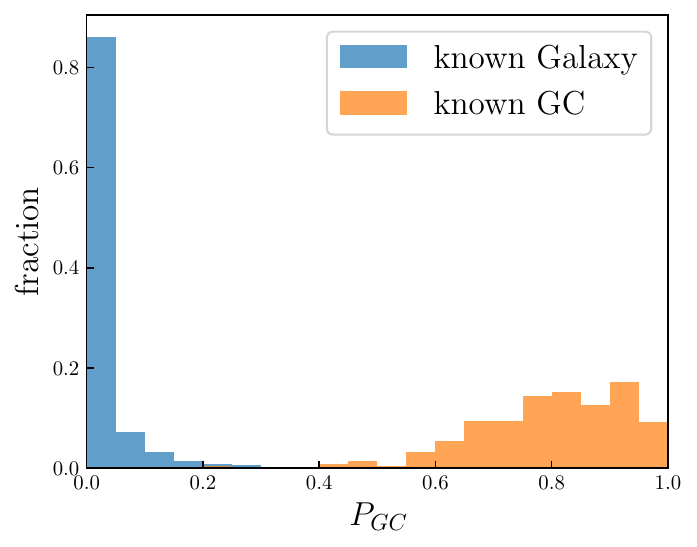}
\caption{\label{fig:pgc_distribution}The distribution of the $P_{GC}$ of the galaxies and GCs with a \textit{simbad} type in our extended sources ($P_{ext}>0.5$). The blue histogram shows the fraction of known galaxies, while the orange histogram shows the fraction of known GCs.}
\end{figure}

Figure~\ref{fig:gdiff_dis} presents the distribution of $G_{pre}{-}G$, which has a median of $\numpm{0.005}{0.022}{0.019}$ and a similar precision as the color $G{-}r$ listed in Table~\ref{tab:lamost_color}. However, $G_{pre}{-}G$ utilizes all four Pan-STARRS colors and the Gaia $G$ magnitude, making it more important than $G{-}r$ in both Clf1 and Clf2. From Table~\ref{tab:parameter importance}, we found that the Clf1 relied more on the Gaia-related photometric parameters [$\log(bp\_rp\_excess)$, $G_{pre}{-}G$, $G{-}r$] and astrometric parameters [$\log(pos\_err)$, $\log(astrometric\_excess\_noise)$]. In contrast to Clf1, the Clf2 did not exhibit a significant reliance on particular parameters. Only the Pan-STARRS colors $z{-}y$ and $i{-}z$ got larger importance, which is 0.101 and 0.095, respectively.
We used Clf1 to obtain the $P_{ext}$ for 1.85 million sources, of which 20658 sources had  $P_{ext}{>}0.5$ and were identified as extended sources. Subsequently, we conducted an inner cross-validation in the training sample for 50 iterations. Each time we randomly split the training sample into two parts, with 80\% of the sources used to train the RF classifier, and the remaining 20\% used to test the accuracy. We evaluated Clf1 using the precision ($P$), the recall ($R$), and the $F_1$ score. These 3 parameters are defined as follows,
\begin{eqnarray}
P&=&\mathrm{\frac{{TP}}{TP+FP}}\label{eq:precision},\\
R&=&\mathrm{\frac{TP}{TP+FN}}\label{eq:recall},\\
F_1&=&2\cdot\frac{P\cdot R}{P+R}\label{eq:F1}.
\end{eqnarray}
Here, we define true positive (TP) as a source that is predicted to be positive and is originally positive, false positive (FP) as a negative source that is predicted to be positive, and false negative (FN) as a positive source that is predicted to be negative. A higher precision indicates lower contamination, while a higher recall indicates more completeness. $F_1$ score is a measure of accuracy, which is the harmonic mean of the precision and the recall, and ranges from 0 to 1, with a higher score indicating better performance. In the inner validation, Clf1 achieved a mean precision=0.981, a mean recall=0.964, and a mean $F_1=0.978$.
With the second classifier Clf2, the $P_{GC}$ is calculated for the extended sources selected by Clf1. The results for sources with a definite \textit{simbad} were plotted in \ref{fig:pgc_distribution}, which showed that about 98.5\% of known galaxies had a $P_{GC}$ lower than 0.25, while the $P_{GC}$ was greater than 0.25 for 99.3\% of known GCs. A catalog of extended sources with the probabilities of $P_{ext}$ and $P_{GC}$ is available in the online version of this manuscript. 

\subsection{Parameter distribution for point and extended sources}\label{section:parameter_distribution}
\begin{figure}[h!]
\centering
\includegraphics[scale=0.73]{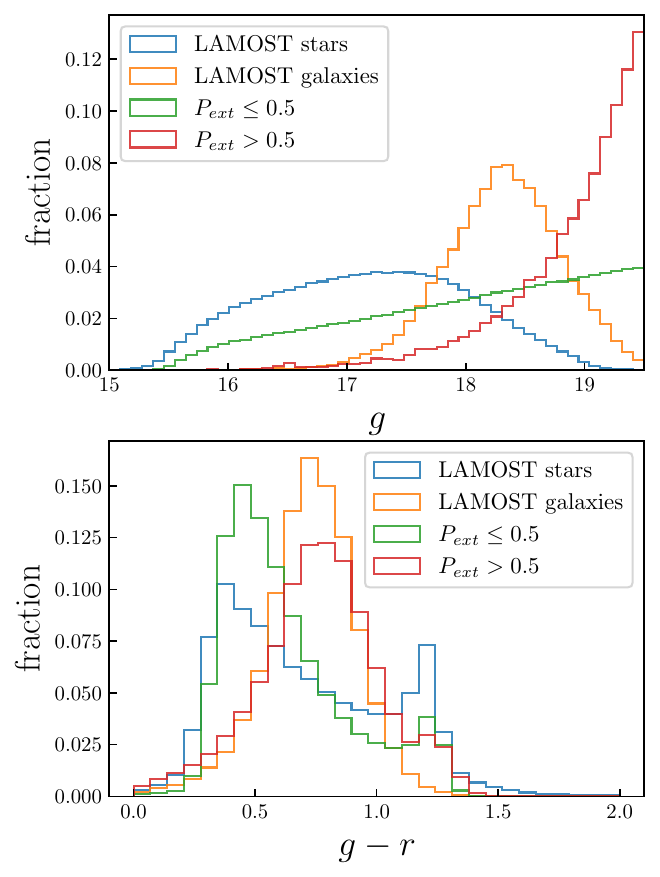}
\caption{ \label{fig:sample_basic_dis}\textbf{Top}: The Pan-STARRS $g$ magnitude distributions of the different samples mentioned in Section~\ref{section:parameter_distribution}. The blue and orange lines represent the LAMOST stars and LAMOST galaxies, respectively. The green and red lines represent the predicted point sources ($P_{ext}{<}0.5$) and extended sources ($P_{ext}{>}0.5$). \textbf{Bottom}: The color $g{-}r$ distributions of the same 4 different samples in the top panel.}
\end{figure}

We compared the distributions of four parameters, namely $\log(bp\_rp\_excess)$, $G_{pre}{-}G$, $\log(pos\_err)$, and $\log(astrometric\_excess\_noise)$, in two different samples. The first sample consists of stars and galaxies from LAMOST DR8 LRS with $SNR{>}20$, while the second sample is composed of our predicted sources, including point sources ($P_{ext}{\leqslant}0.5$) and extended sources ($P_{ext}{>}0.5$). 
The magnitude and color distributions of the two samples are shown in Figure\,\ref{fig:sample_basic_dis}. In the top panel, we observe that most of the LAMOST sources are distributed within $15.5{\sim}18.5\unit{mag}$, while the predicted sources are concentrated at the faint end. The color distributions of the two samples are similar, as shown in the bottom panel.

\begin{figure*}
\centering
\includegraphics[scale=0.69]{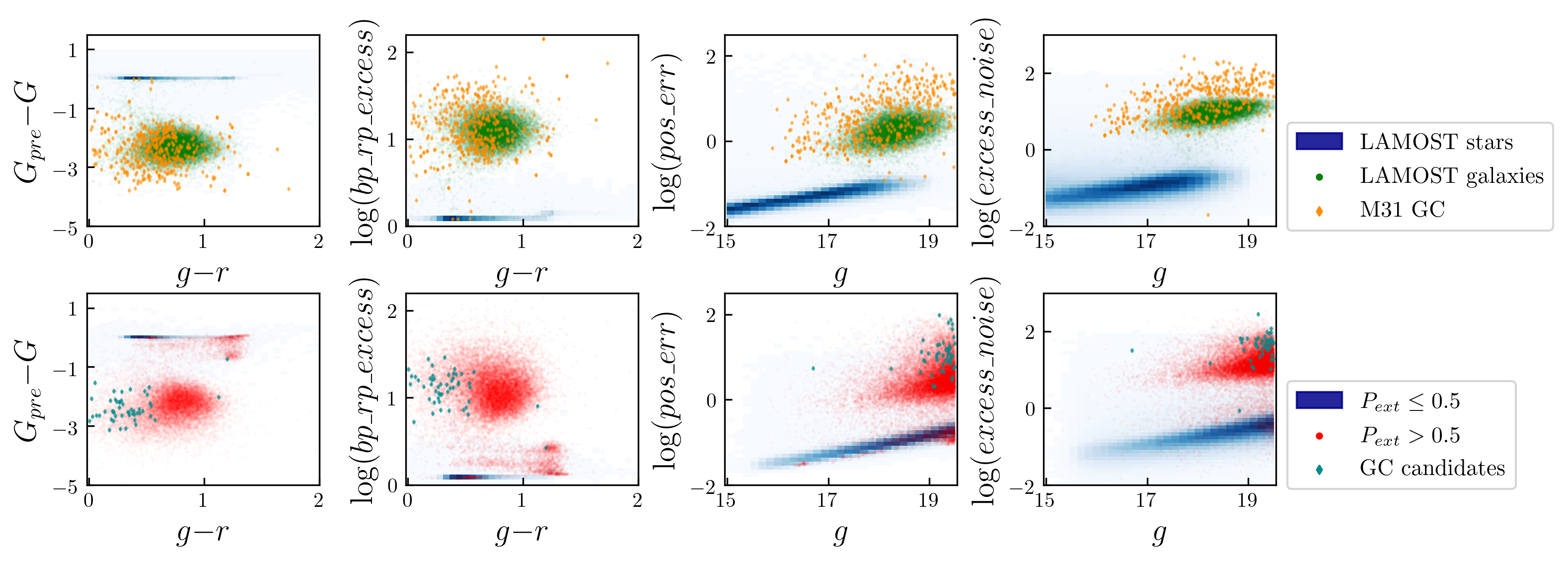}
\caption{\label{fig:extend_point_dis}The distributions of 4 parameters, including  $G_{pre}{-}G$, $\log(bp\_rp\_excess)$, $\log(pos\_err)$ and $\log(astrometric\_excess\_noise)$, which is writen as $\log(excess\_noise)$ in the fourth column. In the \textbf{Top} panel, we plot the LAMOST stars by a 2-dimension histogram, while the LAMOST galaxies are plotted with green dots and the M\,31 GCs with orange diamonds. In the \textbf{bottom} panel, we plot the predicted point sources ($P_{ext}{<}0.5$) in the 2-dimension histogram and the predicted extended sources ($P_{ext}{>}0.5$) in red dots. We also plot the GC candidates in darkcyan diamonds.}
\end{figure*}

We selected four parameters with high $imp_1$ in Table\,\ref{tab:parameter importance}: $G_{pre}{-}G$, $\log(bp\_rp\_excess)$, $\log(pos\_err)$, and $\log(astrometric\_excess\_noise)$, because they are closely related to $G$-band photometry or Gaia astrometry. Their distributions are shown in Figure\,\ref{fig:extend_point_dis}. For the photometry-related parameters, $G_{pre}{-}G$ and $\log(bp\_rp\_excess)$, we plotted them against the color $g{-}r$. Both point sources and extended sources concentrate at different values along the y-axis. For the astrometric parameters $\log(pos\_err)$ and $\log(astrometric\_excess\_noise)$, we plotted them with magnitude $g$, which shows an increasing trend at the faint end, while the point sources and extended sources are still well separated. 

\begin{figure}[h!]
\centering
\includegraphics[scale=0.65]{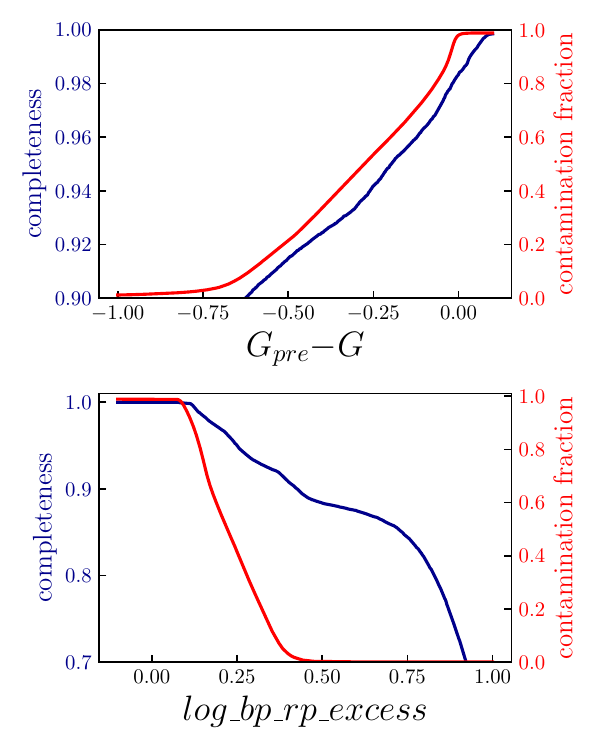}
\caption{\label{fig:cut_predict}\textbf{Top}: The contamination fraction varies with the cut of $G_{pre}{-}G$ in the predicted sample. The blue line shows the completeness with left y-axis, while the red line shows the contamination fraction with the right y-axis. \textbf{Bottom}:The completeness of the extended sources varies with the cut of $\log(bp\_rp\_excess)$. The blue line and red line have the same meaning in the top panel.}
\end{figure}

We aimed to provide practical and effective parameter cuts to facilitate the selection of extended sources, while maintaining low contamination and high completeness in the resulting sample. The performance of the cuts on the $G_{pre}{-}G$ and $\log(bp\_rp\_excess)$ parameters is illustrated in Figure~\ref{fig:cut_predict}, which shows that both the contamination fraction and completeness of the extended sources vary with the cuts. In Tables \ref{tab:cuts_predicted} and \ref{tab:cuts_lamost}, we provide the cuts for these two parameters at different contamination fractions, along with the corresponding completeness values. We also provide the criteria for $\log(pos\_err)$ and $\log(astrometric\_excess\_noise)$ in Table~\ref{tab:cuts_other}, which is a linear line to distinguish point and extended sources. 
The differences in the cuts and performance of the same parameter for the two samples are mainly due to their different source distributions shown in Figure~\ref{fig:sample_basic_dis}. The completeness of the LAMOST sample is consistently better than that of the predicted sample, as shown in Tables~\ref{tab:cuts_predicted} and \ref{tab:cuts_lamost}. We propose 2 reasons to explain it. Firstly, the predicted sample was classified in Sect.\ref{section:classify1}, which only achieved a mean completeness of 0.964. Therefore, the parameter cuts of the predicted sample cannot exceed this completeness. Secondly, the sources in the LAMOST sample are brighter than those in the predicted sample, leading to better performance. It is worth noting that the sample distribution and composition can significantly influence the effectiveness of parameter cuts or machine-learning classifiers. This topic has been discussed in depth in \cite{bailer-jones-qso-galaxy} and \cite{gaia_dr3_non_stellar}. 
Therefore, we suggest using the cuts in Table~\ref{tab:cuts_predicted} and Table~\ref{tab:cuts_lamost}, depending on the $g$ and $g{-}r$ distribution of the candidate sample. Our total sample was selected using $g{-}r{>}0$ and $16{<}g{<}19.5$, which implicitly restricts the use of these parameter cuts. For samples selected from Gaia EDR3, with a similar $g$ and $g{-}r$ distribution as our predicted sample, we recommend using the cuts in Table~\ref{tab:cuts_predicted}. For brighter samples, we recommend using the cuts in Table~\ref{tab:cuts_lamost}. However, note that the completeness and contamination fraction would depend on 
sky positions. For instance, our parameter cuts may generate more contamination in the selected extended sources in the dense area of the Milky Way disk.

\begin{deluxetable}{cccc}[ht]
\tablecaption{Parameter cuts for extended sources in the predicted sample \label{tab:cuts_predicted}}
\tablewidth{0pt}
\tablehead{
\colhead{Parameter} &\colhead{contamination}&\colhead{completeness}
&\colhead{criterion}}
\startdata
 $G_{pre}{-}G$ & $10\%$ & $90.2\%$ & ${<}\, {-}0.613$\\
 $G_{pre}{-}G$ & $20\%$ & $91.3\%$ & ${<} \,{-}0.514$\\
 $G_{pre}{-}G$ & $30\%$ & $92.2\%$ & ${<}\, {-}0.428$\\
 $G_{pre}{-}G$ & $50\%$ & $93.7\%$ & ${<} \,{-}0.278$\\
 $\log\,(bp\_rp\_excess)$ & $10\%$ & $92.1\%$ & ${>}$ 0.360\\
  $\log\,(bp\_rp\_excess)$ & $20\%$ & $92.8\%$ & ${>}$ 0.322\\
 $\log\,(bp\_rp\_excess)$ & $30\%$ & $93.6\%$ & ${>}$ 0.288\\
 $\log\,(bp\_rp\_excess)$ & $50\%$ & $96.2\%$ & ${>}$ 0.221\\
\enddata 
\tablecomments{Col2 is the contamination fraction of the extended sources selected by the criterion in Col4. Col3 is the completeness of all the extended sources in the predicted sample. For most samples selected from Gaia EDR3, with a magnitude of $g{\lesssim}19.5$, we recommend using the cuts of $\log(bp\_rp\_excess)$, which is available for every source in Gaia. With a regression of predicted $G$-band magnitude, it is also feasible to use the cuts for $G_{pre}{-}G$.}
\end{deluxetable}
\begin{deluxetable}{cccc}[ht]
\tablecaption{Parameter cuts for extended sources in the LAMOST sample\label{tab:cuts_lamost}}
\tablewidth{0pt}
\tablehead{
\colhead{Parameter} &\colhead{contamination}&\colhead{completeness}
&\colhead{criterion}}
\startdata
 $G_{pre}{-}G$ & $10\%$ & $99.0\%$ & ${<}\,{-}0.650$\\
 $G_{pre}{-}G$ & $20\%$ & $99.1\%$ & ${<}\,{-}0.514$\\
 $G_{pre}{-}G$ & $30\%$ & $99.2\%$ & ${<}\,{-}0.428$\\
 $G_{pre}{-}G$ & $50\%$ & $99.3\%$ & ${<}\,{-}0.278$\\
 $\log\,(bp\_rp\_excess)$ & $10\%$ & $99.1\%$ & ${>}\,0.378$\\
  $\log\,(bp\_rp\_excess)$ & $20\%$ & $99.1\%$ & ${>}\,0.329$\\
 $\log\,(bp\_rp\_excess)$ & $30\%$ & $99.2\%$ & ${>}\,0.286$\\
 $\log\,(bp\_rp\_excess)$ & $50\%$ & $99.3\%$ & ${>}\,0.214$\\
\enddata 
\tablecomments{Columns 1-4 have the same meaning as Table~\ref{tab:cuts_predicted}. These cuts can be used for samples that have similar distribution as our LAMOST sample, which is shown in Figure~\ref{fig:sample_basic_dis}.}
\end{deluxetable}
\begin{deluxetable*}{ccccc}[ht!]
\tablecaption{Parameter cuts for two astrometric parameters}\label{tab:cuts_other}
\tablewidth{0pt}
\tablehead{
\colhead{sample} & \colhead{parameter} & \colhead{  contamination  } & \colhead{  completeness  } & \colhead{criteria}
}
\startdata
 Predicted & $\log\,(astrometric\_excess\_noise)$ & $30\%$ & $84.0\%$ & ${>}\,0.050{\cdot}g{-}0.206$\\
 Predicted & $\log\,(astrometric\_excess\_noise)$ & $50\%$ & $88.5\%$ & ${>}\,0.031{\cdot} g{-}0.207$\\
 Predicted & $\log\,(pos\_err)$ & $30\%$ & $75.9\%$ & ${>}\,0.027{\cdot}g{-}0.414$\\
 Predicted & $\log\,(pos\_err)$ & $50\%$ & $87.8\%$ & ${>}\,0.007{\cdot} g{-}0.413$\\
 LAMOST& $\log\,(astrometric\_excess\_noise)$ & $30\%$ & $99.0\%$ & ${>}\,0.040{\cdot}g{-}0.517$\\
 LAMOST & $\log\,(astrometric\_excess\_noise)$ & $50\%$ & $99.6\%$ & ${>}\,0.056{\cdot} g{-}1.345$\\
 LAMOST & $\log\,(pos\_err)$ & $30\%$ & $68.0\%$ & ${>}\,0.014{\cdot}g{-}0.138$\\
 LAMOST & $\log\,(pos\_err)$ & $50\%$ & $97.8\%$ & ${>}\,{-}0.413$
\enddata
\tablecomments{The linear parameter cuts for the astrometric parameters, $\log\,(astrometric\_excess\_noise)$ and $\log\,(pos\_err)$. Col1 points the sample used to get the criteria. Col 2-5 have the same meaning as Col 1-4 in Table~\ref{tab:cuts_predicted} and \ref{tab:cuts_lamost}. These cuts can also be used to select extended sources, but we recommend using the cuts in Table~\ref{tab:cuts_other} and Table~\ref{tab:cuts_lamost}, considering the contamination fraction and completeness. }
\end{deluxetable*} 
\subsection{Confused sources} 
\label{section:result_point_extend}
\begin{figure}[h!]
\centering
\includegraphics[scale=0.66]{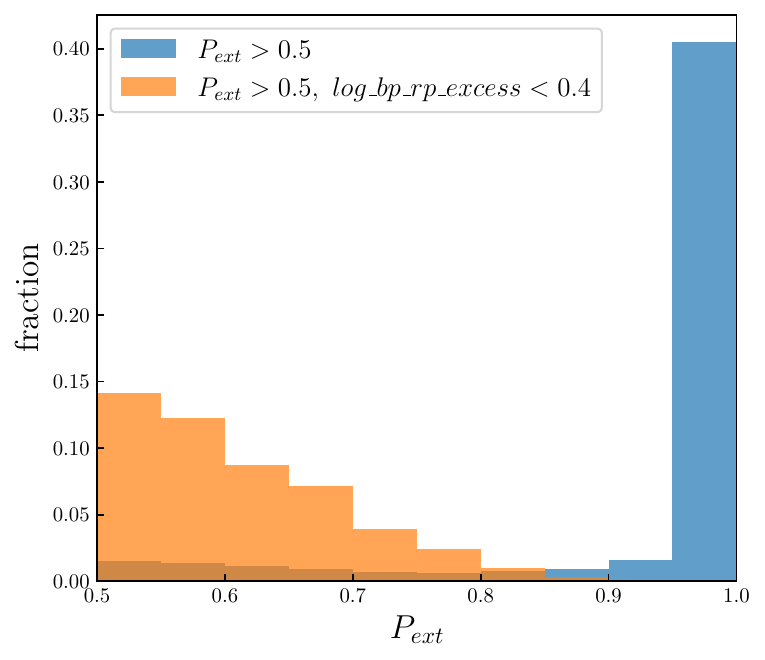}
\caption{ \label{fig:ss2_sp_y1p}The blue histogram is the $P_{ext}$ distribution of the predicted extended sources ($P_{ext}{>}0.5$). The orange histogram is a small part of the predicted extended sources, with $log\_bp\_rp\_excess{<}0.4$ and $P_{ext}{>}0.5$. }
\end{figure}

In Figure~\ref{fig:extend_point_dis}, we notice some extended sources fall in the range of point sources. We discuss this phenomenon for the LAMOST sample and the predicted sample, respectively.
First, we applied our Clf1 to the LAMOST sample and get their $P_{ext}$. Out of 26,778 LAMOST galaxies, 268 had $P_{ext}$ values below 0.5, while out of 1.7 million LAMOST stars, 578 had $P_{ext}$ values above 0.5. We cross-matched these sources with \textit{simbad}, and only part of them got a result of the \textit{simbad} types, which are listed in Table~\ref{tab:wrong_tag_sources}. Among the galaxies with $P_{ext}{<}0.5$, the main types were galaxies and QSOs, while among the stars with $P_{ext}{>}0.5$, the main types were galaxies and GCs. Since our $P_{ext}$ is generated from the Clf1 based on RBC V.5 and \textit{simbad}, the $P_{ext}$ has higher consistency with the \textit{simbad} type.
In Figure~\ref{fig:extend_point_dis}, we observe an additional concentration of extended sources at the point source region in the bottom panel of the predicted sample. Specifically, around 1,900 to 2,000 out of 20,658 extended sources have $log\_bp\_rp\_excess{<}0.4$ or $G_{pre}{-}G{>}{-}0.6$. Among these sources, 50 have a known \textit{simbad} main type, consisting of 26 stars, 17 galaxies, 15 QSOs, and 2 other types, while 54 have a LAMOST class, comprising 50 stars and 4 QSOs. In Figure~\ref{fig:ss2_sp_y1p}, the distribution of the sources ($log\_bp\_rp\_excess{<}0.4$ and $P_{ext}{>}0.5$) is distinct from that of the extended sources ($P_{ext}{>}0.5$). We believe that these sources are predominantly composed of a small fraction of extended sources and a large fraction of inevitable contamination from point sources. This is reasonable considering the precision of our point-extended classifier Clf1 in Section~\ref{section:classify1} and the characteristics of our training sample in Section~\ref{section:training_sample}. Since these sources ($P_{ext}{>}0.5$, $log\_bp\_rp\_excess{<}0.4$ or $G_{pre}{-}G{>}{-}0.6$) are classified as extended sources by Clf1 but rejected as point sources by the parameter cuts in Table~\ref{tab:cuts_other}, it indicates that our cuts in Table~\ref{tab:cuts_other} can yield a sample with fewer false positives (FP) than anticipated, equivalent to less contamination than the value listed in the same table.
\begin{deluxetable}{ccc}[h!]
\tablecaption{Simbad main type of LAMOST stars and galaxies with a contradict $P_{ext}$\label{tab:wrong_tag_sources}}
\tablewidth{0pt}
\tablehead{
\colhead{ $\ \ \ \ \ $  \textit{simbad} $ \ \ \  $  } &\colhead{ $ \ \ \  $  Stars $ \ \ \  $}&\colhead{ $ \ \ \  $ Galaxies  $\ \ \ \ \ $ }\\
\colhead{Type} &\colhead{$P_{ext}{>}0.5$}&\colhead{$P_{ext}{<}0.5$}}
\startdata
total\tablenotemark{a} & 227 (35154) & 122 (18013) \\
star & 9 & 4\\
galaxy & 131 &53\\
AGN & 5 & 65\\
other & 81 & 0
\enddata 
\tablecomments{Col1 is the \textit{simbad} type, including dozens of types, which is divided  into 4 categories, including star, galaxy, AGN and other. Taking AGN as an example, it contains Seyfert, BL Lac, Blazer, QSO, and so on. Col2 is the LAMOST stars with $P_{ext}{>}0.5$, while Col3 is the LAMOST galaxies with $P_{ext}{<}0.5$. }
\tablenotetext{a}{In the bracket, it is the total number of the LAMOST stars or galaxies that have a \textit{simbad} main type. Outside the bracket, it is the number of sources with contradicting $P_{ext}$ and a \textit{simbad} main type. It means that no more than 1\% of these sources have a contradicted $P_{ext}$. }
\end{deluxetable}
\subsection{Compare extended sources with existed catalogs}\label{section:result_compare_catalogs}
We compared our extended sample with three existing catalogs of the M\,31 clusters: RBC V5, AP cluster catalog, and NVK catalog of star cluster candidates \citep{NVK2008}. The RBC V5 catalog is compiled from historical data with further improvements from multiple observations. The AP and NVK catalogs searched for star clusters from high-resolution images, such as those taken by the HST and the Subaru Telescope. 

In Table~\ref{tab:other_catalogs}, we list the result of our comparison. We found that approximately 77\% of RBC GCs fall within our magnitude and color range, of which 99\% are classified as extended sources ($P_{ext}{>}0.5$). Among the 4 RBC sources with $P_{ext}{<}0.5$, three are confirmed stars \citep{Huxor2014}, while one is a possible GC \citep{ugrizK_2010}. This GC is classified by its color and images of the SDSS and United Kingdom Infrared Telescope (UKIRT) \citep{RBC2004}. However, it shows a similar $G_{pre}{-}G{=}{-}0.013$ and $log\_bp\_rp\_excess{=}0.084$ as stars, which indicated that it is actually a point source. Two sources with $P_{ext}{>}0.5$ have $P_{GC}{<}0.25$. One of these sources corresponds to a star cluster found in the AP catalog, which presents a challenge for our classifiers. This difficulty arises due to the fact that our training samples are primarily based on the GCs in RBC V5 and \textit{simbad}, which are mostly identified with the observations made by ground-based telescopes. In contrast, the AP catalog is generated from high-resolution HST images. Since our training sample includes only a limited number of star clusters from the AP catalog, unless they are also classified as GCs, the classifier Clf2, trained using all GCs, assigns a low $P_{GC}$ value to this particular source. The other source is labeled as a GC in RBC V5, which is labeled as a star in \textit{simbad} due to its radial velocity $RV{=}33\unit{km\cdot s^{-1}}$ \citep{Kang2012}. Therefore, we also labeled it as a star in our training sample in Section~\ref{section:training_sample}, so that it got a low $P_{GC}$.

Our comparison with the AP and NVK catalogs yielded a match of only 30\% of sources that met our requirements for magnitude and color. This is likely due to the majority of sources in these catalogs being fainter than our sample. In the matched sources, about $81\%\sim85\%$ of sources satisfied the same criteria for GCs in the RBC V5, namely $P_{ext}{>}0.5$ and $P_{GC}{>}0.25$. It is worth noting that the HST and Subaru have a higher resolution than the telescopes used in the RBC, enabling the discovery of smaller star clusters that would be ignored in previous GC searching projects. Consequently, 13.7\% of sources in AP and 8.3\% of sources in NVK had $P_{ext}<0.5$, which is a higher ratio than that in the RBC V5 (1\%).
\begin{deluxetable}{cccc}[ht]
\tablecaption{\label{tab:other_catalogs}The cross-match result between our extended sources and 3 existed catalogs.}
\tablewidth{0pt}
\tablehead{
\colhead{Catalog} &\colhead{Sample size\tablenotemark{a}}&\colhead{$P_{ext}{>}0.5$}
&\colhead{$P_{ext}{>}0.5\,\&\,P_{GC}{>}0.25$}}
\startdata
RBC V.5 & $379 \,(77\%)$ & $375$ & 373\\
AP & $194 \,(29\%)$ & $167$ & 159\\
NVK & $48\, (31\%)$ & $44$ & 41
\enddata 
\tablecomments{Col1 is the catalog name. Col2 is the total number of the cross-match result. Col3 and col4 list the number of sources that satisfy our criteria.}
\tablenotetext{a}{The number of sources that fall in the same color and magnitude range as our total sample. The percentage is the fraction of the sample in the origin catalog.}
\end{deluxetable}
\subsection{GC candidates}\label{section:result_candidates}
\begin{figure}[h!]
\centering
\includegraphics[scale=0.185]{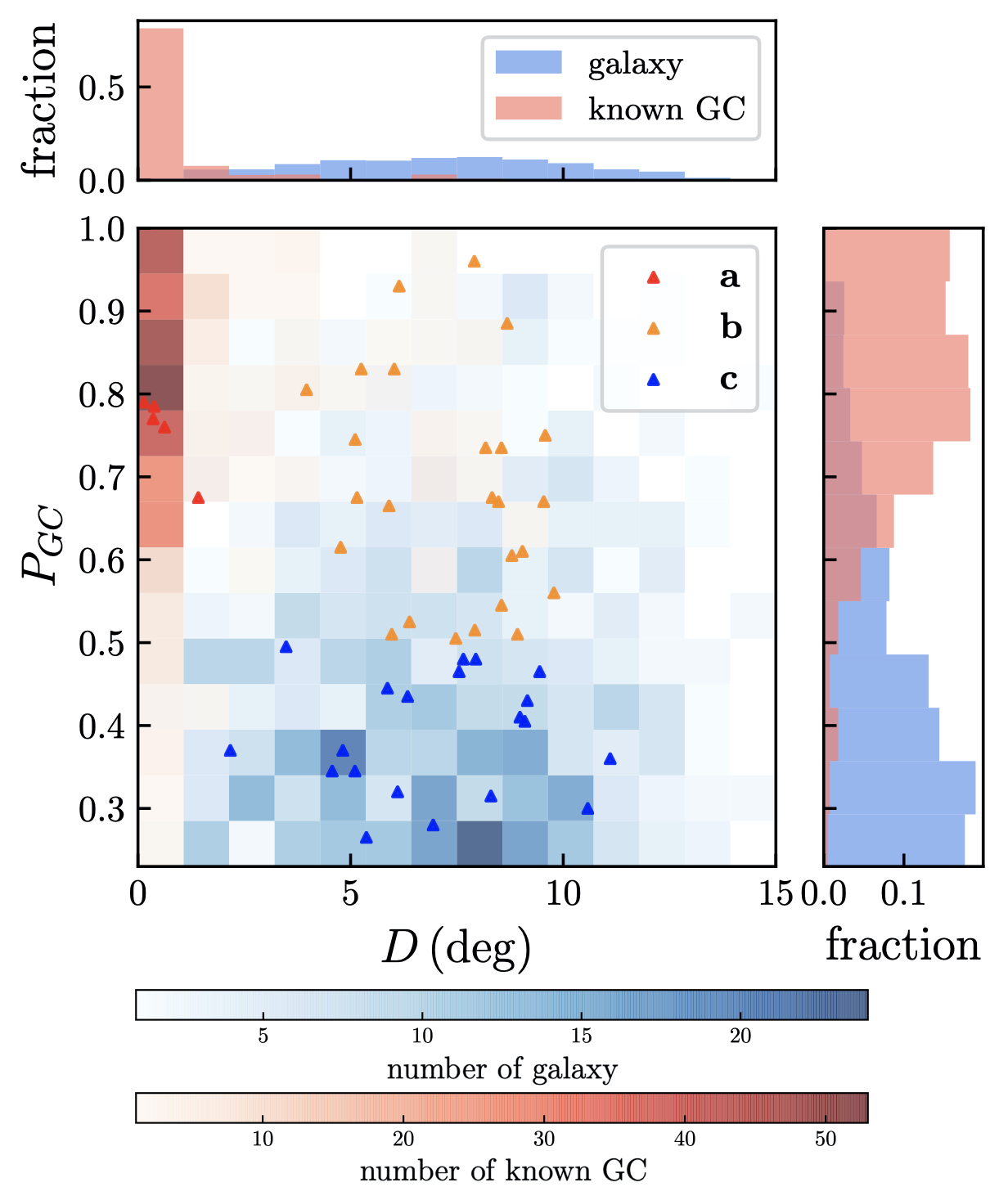}
\caption{\label{fig:candidates_and_known_sources}The candidates, the known GCs and galaxies in our 1562 visually checked sources are plotted in the $P_{GC}-D$ plane. In the middle panel, the triangles stand for the candidates, with the different colors indicate their different types. The red squares show the distribution of the known GCs. The blue squares show the distribution of the known galaxies, which is also confirmed by our visual inspections. All of these sources get a $P_{GC}$ higher than 0.25. We use the visually checked galaxies here, because only 116 \textit{simbad} labeled galaxies have $P_{GC}{>}0.25$, even though we obtained over 4000 of them in total with $P_{GC}{<}0.25$. Thus, the galaxies, used in this figure, are the galaxies that are most likely recognized as GCs by our Clf2. The top panel is the distance distribution of the known GCs and galaxies, while the right panel is distribution of $P_{GC}$. The height of the histogram means the fraction of the sample in each bin. The color of the bars in both top and right panels refers to the same sample as the squares in the middle panel.}
\end{figure}

After the visual inspection, we got 50 new GC candidates, excluding the known GCs and the star clusters in AP catalog. Source 1 in Section~\ref{section:visual_inspec} is kept since it is a GC candidate in RBC V5 and a stellar cluster candidate in NVK catalog. Their basic information are listed in Table~\ref{tab:candidates50}. We found no overlap between our candidate list and that of \cite{WSC_2022} since their candidates are fainter than ours. The  GC candidates were divided into 3 types, from type \textbf{a} to type \textbf{c}, according to their $P_{GC}$ and the projected distance $D(\rm{deg})$ to the center of M\,31. The criteria are listed in Table~\ref{tab:candidates_number}. In Figure~\ref{fig:candidates_and_known_sources}, we show the distribution of these candidates in a $P_{GC}{-}D$ plane, along with the distributions of 818 galaxies confirmed by our visual inspection and 410 known GCs.
\begin{deluxetable}{cccc}[ht]
\tablecaption{\label{tab:candidates_number}Three types of GC candidates.}
\tablewidth{0pt}
\tablehead{
\colhead{$ \ \ \ $Type $ \ \ \  $} &\colhead{$ \ \ \  $$D\, (\rm{deg})$$ \ \ \  $}&\colhead{$ \ \ \  $$P_{GC}$$ \ \ \  $}
&\colhead{$ \ \ \  $number$ \ \ \  $}}
\startdata
$\mathbf{a}$ & ${\leqslant}2$ & $>\!0.5$ & 5 \\
$\mathbf{b}$ & $>\!2$ & $>\!0.5$ & 25 \\
$\mathbf{c}$ & $>\!2$ & $\leqslant\!0.5$ & 20 
\enddata 
\tablecomments{We divided the  candidates into 3 types, according to the criteria in col2 and col3, which is the projected distance (in degree) to the center and $P_{GC}$. }
\end{deluxetable}

In Figure~\ref{fig:candidates_and_known_sources}, approximately $85\%$ of the GCs concentrate within the region of $[D{<}2^{\circ},\ P_{GC}{>}0.5]$,  while around $62\%$ of the galaxies ($P_{GC}{>}0.25$) distribute in the area of $[D{>}2^{\circ},\ P_{GC}{<}0.5]$. The two categories overlap at the region of $[D{>}2^{\circ},\ P_{GC}{>}0.5]$, which includes $9.5\%$ GCs and $32.6\%$ galaxies. The 6 type-$\mathbf{a}$ candidates are the most likely candidates. We got 25-type $\mathbf{b}$ candidates in the overlapped area, and 20-type $\mathbf{c}$ candidates in the area dominated by galaxies. Given that we have already eliminated as many elliptical galaxies as possible, we propose that these 45 remaining candidates warrant a spectroscopic study, with particular attention given to the 25 type-$\mathbf{b}$ candidates. The PAndAS images of all the  candidates are shown in Figure~\ref{fig:candidates_figure_50}. 
\begin{figure}[h!]
\centering
\includegraphics[scale=0.71]{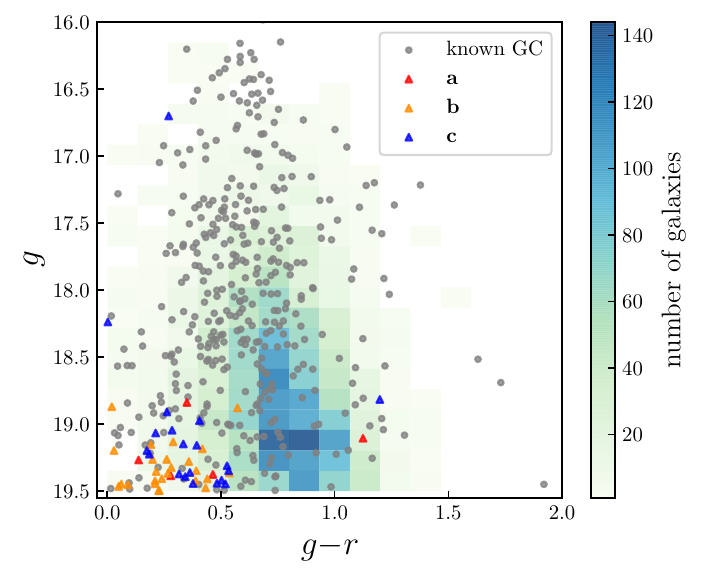}
\caption{\label{fig:candidate_color_magnitude}
The distribution of the  candidates (triangles) in the color-magnitude plane, together with the known GCs and the known galaxies. The 2-D histogram shows the distribution of known galaxies, which are labeled as Galaxy in \textit{simbad} with $P_{ext}{>}0.5$. The grey dots are the known GCs.}
\end{figure}

In the bottom panel of Figure~\ref{fig:extend_point_dis}, we plot the GC candidates with darkcyan dots, all of which fall in the range of extended sources. Compared to the top panel, these candidates show a similar distribution as the known GCs (red dots in the top panel) for all 4 parameters (along the y-axis). In Figure~\ref{fig:candidate_color_magnitude}, the magnitude and color of the candidates are shown with the known GCs and galaxies ($P_{ext}{>}0.5$ and labeled as Galaxy in \textit{simbad}). The candidates appear fainter compared to the known GCs, as previous studies have likely already identified most of the brighter GCs. Our candidates predominantly occupy the fainter end of the distribution, with $g{-}r{\lesssim}0.5$, avoiding the region densely populated by galaxies. We attribute this clustering as a selection effect in our classification process, where the abundance of galaxies in this region contributes to a lower $P_{GC}$ assigned to potential GC candidates. As only sources with $P_{GC}{>}0.25$ underwent visual inspection, potential candidates in this particular region were missed. Consequently, our candidates predominantly appear in the bluer region. Furthermore, there is no difference in magnitude and color between different types of candidates. 

\begin{figure}[ht!]
\centering
\includegraphics[scale=0.76]{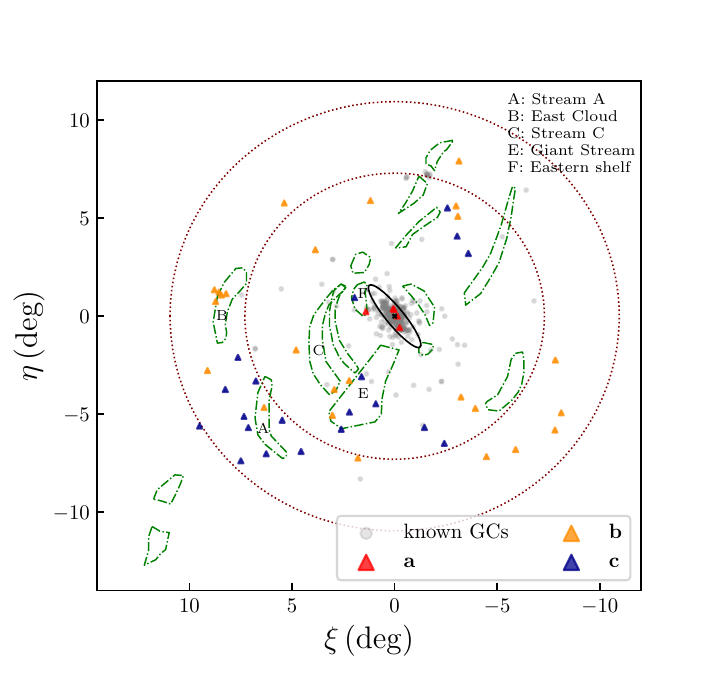}
\caption{\label{fig:spatial_dis}The spatial distribution of the  candidates. Known GCs are depicted as grey dots, while triangles represent different types of candidates. The dark red dotted line shows circles with radius of $100\unit{kpc}$ and $150\unit{kpc}$, centered at coordinate (0, 0), which stand for the center of M\,31. The black ellipse indicates the disk of M\,31, also centered at (0, 0), with a major axis of $2^\circ$. Here, we adopt a major axis position angle of $38^\circ$, with an inclination angle of $78^\circ$ \citep{wangsong2019}. The irregular green dash-dotted lines show the large-scale structures of M\,31 (copied from Figure 12 of \cite{pandas_structure_2018}). Some GC candidates fall in several structures, which are listed in Table\,\ref{tab:candidates_in_structures}. The names of the structures are indicated in the upper right corner of the figure.}
\end{figure}

\begin{figure*}[h!]
\centering
\includegraphics[scale=0.65]{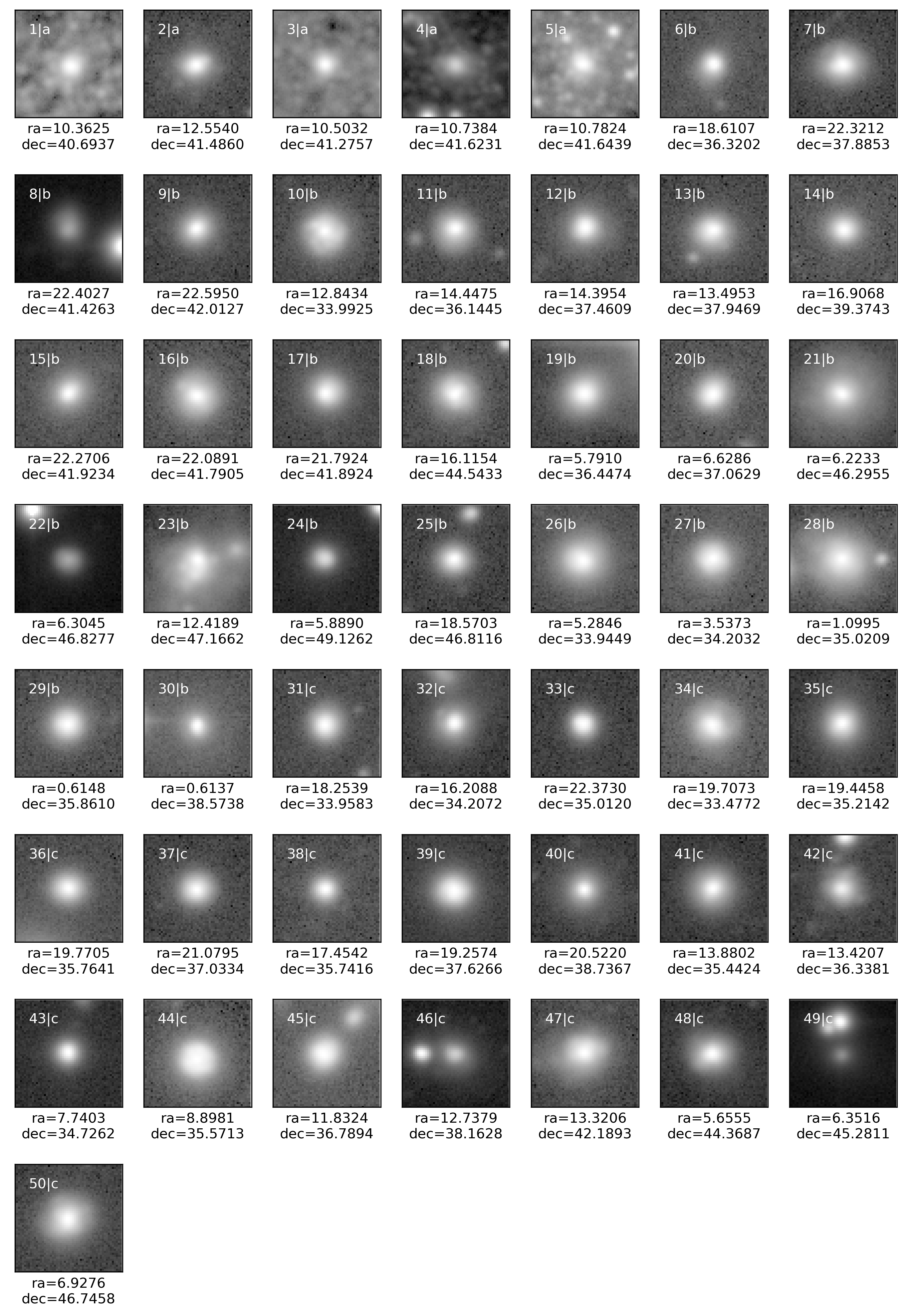}
\caption{\label{fig:candidates_figure_50} PAndAS $g$-band images of the 50 candidates newly identified in this work. The size of each panel is $54{\times}54$ pixels, which is $10.1{\times}10.1\unit{arcsec^2}$.  On the top left in each panel, we wrote the ID and type of this source. The position (J2000) of the source is at the bottom of the panel.}
\end{figure*}

The spatial distribution of our candidates is illustrated in Figure~\ref{fig:spatial_dis}. A majority of the candidates are situated outside the disk of M\,31, with 25 of them extending beyond a distance of $100\unit{kpc}$. The farthest candidate is located $152\unit{kpc}$ away from the center of M\,31. This projected distance is consistent with previous findings in the literature. For instance, \cite{Huxor2014} identified an M\,31 GC at a distance of 141.34\,kpc, while the most distant M\,31 GC candidate discovered by \cite{WSC_2022} was found at a projected distance of 158\,kpc. In our own Milky Way, \cite{milkyway_farthest_gc} detected a GC at a distance of $147\pm17\unit{kpc}$. Additionally, \cite{ngc5128_gc} observed GCs in NGC\,5128 and identified one at a projected distance of 187\,kpc. In the case of M\,81, M\,81-GC2 was observed to extend up to a distance of 400\,kpc along our line of sight \citep{M81_GC2}. Figure~\ref{fig:spatial_dis} displays 14 GC candidates associated (in projection) with the large-scale structures in the galaxy halo of M\,31 \citep{pandas_structure_2018}, including Stream A, the East Cloud, Stream C, the Giant Stream, and the Eastern shelf. The coordinates and corresponding structure labels for these candidates are listed in Table~\ref{tab:candidates_in_structures}. 

\begin{deluxetable}{cccccc}[ht]
\tablecaption{\label{tab:candidates_in_structures}The GC candidates fall in the large-scale structures of M\,31.}
\tablewidth{0pt}
\tablehead{
\colhead{Structure} &\colhead{$\xi$}&\colhead{$\eta$}
&\colhead{R.A.}&\colhead{Dec}&\colhead{Type}}
\startdata
A&6.37&-4.65&18.61068&36.32024&b\\
B&8.56&1.23&22.27058&41.92345&b\\
B&8.45&1.08&22.08908&41.79052&b\\
B&8.22&1.15&21.79239&41.8924&b\\
B&8.73&0.75&22.40269&41.42634&b\\
C&2.94&-3.74&14.3954&37.46088&b\\
E&2.22&-3.28&13.49527&37.94687&b\\
E&1.61&-3.09&12.7379&38.16275&c\\
E&2.2&-4.89&13.42074&36.33812&c\\
E&3.04&-5.05&14.44746&36.1445&b\\
E&2.6&-5.77&13.88016&35.44242&c\\
E&0.92&-4.47&11.83242&36.78944&c\\
F&1.95&0.95&13.32059&42.18934&c\\
F&1.4&0.23&12.554&41.48601&a
\enddata 
\tablecomments{Col1 is the same label of the structures as Figure\,\ref{fig:spatial_dis}, A for the Stream A, B for the East Cloud, C for the Stream C, E for the Giant Stream and F for the Eastern shelf. Col2 and Col3 are the coordinates used in Figure\,\ref{fig:spatial_dis}. Col4 and Col5 is the celestial coordinates of epoch J2000. Col6 is the type of the candidates.}
\end{deluxetable}

\begin{deluxetable*}{cccccccccccc}[ht!]
\tablecaption{Catalog of the visually checked GC candidates}\label{tab:candidates50}
\tablewidth{0pt}
\tablehead{
\colhead{ID} & \colhead{Type} & \colhead{RA} & \colhead{DEC} & \colhead{$G$}&  \colhead{$g$} & \colhead{$g{-}r$} & \colhead{$\log\,(bp\_rp\_excess)$} &\colhead{$G_{pre}-G$} &\colhead{$P_{ext}$}& \colhead{$P_{GC}$}
}
\startdata
1&a&00:41:27.00&+40:41:37.32&21.22&19.39&0.28&1.04&-2.15&1.0&0.76&\\
2&a&00:50:12.96&+41:29:09.60&21.47&19.27&0.14&1.13&-2.52&0.96&0.68\\
3&a&00:42:00.77&+41:16:32.52&19.75&19.11&1.12&0.9&-2.03&1.0&0.79\\
4&a&00:42:57.22&+41:37:23.16&21.29&18.84&0.35&1.27&-2.63&1.0&0.77\\
5&a&00:43:07.78&+41:38:38.04&21.66&19.38&0.46&1.27&-2.49&1.0&0.78\\
6&b&01:14:26.57&+36:19:12.72&20.78&19.46&0.05&0.72&-1.54&0.76&0.96\\
7&b&01:29:17.09&+37:53:07.08&21.64&19.2&0.03&1.25&-2.69&0.99&0.75\\
8&b&01:29:36.65&+41:25:34.68&21.88&19.42&0.21&1.37&-2.84&1.0&0.6\\
9&b&01:30:22.80&+42:00:45.72&21.08&19.26&0.26&1.0&-2.19&0.96&0.51\\
10&b&00:51:22.42&+33:59:33.00&21.7&19.15&0.19&1.22&-2.79&1.0&0.5\\
11&b&00:57:47.40&+36:08:40.20&21.41&19.26&0.2&1.15&-2.5&0.99&0.66\\
12&b&00:57:34.90&+37:27:39.24&21.35&19.48&0.43&1.16&-2.43&1.0&0.62\\
13&b&00:53:58.87&+37:56:48.84&21.23&19.37&0.27&1.05&-2.2&1.0&0.8\\
14&b&01:07:37.63&+39:22:27.48&21.18&19.41&0.24&1.07&-2.18&1.0&0.74\\
...&...&...&...&...&...&...&...&...&...&...
\enddata
\tablecomments{The ID is the same as Figure~\ref{fig:candidates_figure_50}, while the Type is the same as Table~\ref{tab:candidates_number}. RA and DEC are the J2000 position of the source. $G$ is the Gaia $G$-band magnitude, while $g$ and $g{-}r$ are the magnitude and color of Pan-STARRs. $\log(bp\_rp\_excess)$ is the logarithm of the Gaia parameter $bp\_rp\_excess$, described in Section~\ref{section:gaia_data}. $G_{pre}{-}G$ is the difference between the $G$-band magnitude predicted in Section~\ref{section:regression} and the real Gaia $G$ magnitude. $P_{ext}$ shows the likelihood of a source to be extended, while $P_{GC}$ shows the probability to be a GC. For detail about these 2 probability, please refer to Sections~\ref{section:classify1} and \ref{section:classify2}. The entire catalog is available in the online version of this manuscript.}
\end{deluxetable*} 
\begin{deluxetable*}{ccccccccccc}[ht!]
\tablecaption{Catalog of the 20,658 extended sources classified by Clf1}\label{tab:candidates_extend}
\tablewidth{0pt}
\tablehead{
\colhead{Gaia ID} & \colhead{RA} & \colhead{DEC} & \colhead{$G$}&  \colhead{$g$} & \colhead{$g{-}r$} & \colhead{$\log\,(bp\_rp\_excess)$} &\colhead{$G_{pre}-G$} &\colhead{$P_{ext}$}& \colhead{$P_{GC}$}
}
\startdata
308447268251051136&01:05:41.71&+29:18:43.92&21.91&19.26&0.58&1.43&-3.12&1.0&0.02\\
308447818006910592&01:05:52.46&+29:21:20.16&19.76&18.85&0.93&0.88&-1.74&1.0&0.07\\
308450296202907136&01:05:03.50&+29:22:23.16&20.43&18.91&0.99&1.04&-2.33&1.0&0.0\\
308450708520051328&01:05:20.74&+29:20:58.92&21.37&19.24&0.29&1.09&-2.38&0.92&0.7\\
308451189556387456&01:05:37.34&+29:24:24.84&20.08&18.37&0.93&1.17&-2.54&1.0&0.0\\
308451945470344448&01:05:15.55&+29:25:09.48&20.61&19.39&0.78&0.95&-1.9&1.0&0.09\\
308452392147229696&01:05:15.10&+29:25:58.08&20.18&19.0&0.67&0.84&-1.78&0.92&0.02\\
308455896840256256&01:06:16.87&+29:27:12.24&19.89&19.14&0.95&0.78&-1.62&0.98&0.01\\
308456206077899648&01:06:11.76&+29:29:09.60&19.82&18.74&1.06&1.02&-1.93&1.0&0.01\\
...&...&...&...&...&...&...&...&...&...
\enddata
\tablecomments{Col1 is the source\_ID in Gaia EDR3. The other columns have the same meaning as the columns with the same name in Table~\ref{tab:candidates50}. The entire catalog is available in the online version of this manuscript.}
\end{deluxetable*} 

\section{Summary and future perspective}
\label{section:summary_discussion}
Using the catalog data of Gaia EDR3 and Pan-STARRS1, we trained two Random Forest classifiers to search the GC candidates from 1.85 million sources in the large area around M\,31. Among the classifiers, Clf1 is reliable in distinguishing between point sources and extended sources, while Clf2 is not as accurate in distinguishing between globular clusters and dense galaxies. The sources in the RBC V.5 catalog and \textit{simbad} are selected as the training sample. The first classifier yielded 20,658 extended sources, while the second classifier produced 1,562 initial candidates. We visually inspected the PAndAS images of these candidates and selected 50 GC candidates, including one with a clear cluster appearance in its HST image. These candidates were divided into three types based on their $P_{GC}$ and projected distance from the center of M\,31. Most of these candidates are located in the halo of the M\,31, while 14 of them are associated (in projection) with the large-scale structures in the M\,31 halo. The farthest candidate reaches a distance of $152\unit{kpc}$ from the center of M\,31. Follow-up spectroscopic observations could help further eliminate contamination from extra-galactic sources. 
We also distinguished point and extended sources by using catalog data of Gaia EDR3 and Pan-STARRS1, and provided a series of parameter cuts for convenient classification. All of the sources in the Gaia catalog get the parameters, like $bp\_rp\_excess$ and $astrometric\_excess\_noise$, therefore our parameter cuts can be easily applied to other Gaia samples, with consideration of our suggestions in Sections \ref{section:parameter_distribution} and \ref{section:result_point_extend}. The space-based Gaia mission is an all-sky survey with high instantaneous spatial resolution, thus it enable us to select more extended sources than those ground-based surveys. 
The high-resolution image is always the gold criterion for a GC in M\,31 since the era of Edwin Hubble. With the help of HST images, we excluded some contamination, that cannot be discerned through PAndAS images. However, only a small part of M\,31 is observed by HST.
In the future, more high-resolution images could be provided by the Chinese Space Station Telescope (CSST) and Euclid Telescope, which have larger Field of View (FoV) of $1.1\unit{deg^2}$ and $0.55\unit{deg^2}$ \citep{CSST_OS}, respectively, compared to the $0.186\unit{arcmin^2}$ of HST \citep{HST_FOV}. Therefore, there would be new opportunities to classify these GC candidates and search for new GCs in M\,31, especially those in its outer halo. A larger and cleaner sample of GCs in M\,31 will be valuable in research of the structure, formation history, and many other fields of M\,31. 
In addition, GCs can be used to study the intermediate-mass Black Holes (IMBHs), which are possibly located in their centers. \cite{G1_BH} reported a $10^5\unit{M_{\odot}}$ BH in the most massive GC of M\,31, B023-G078. Except for GCs, an IMBH can also reside in a Hyper-Compact Stellar Cluster (HCSC). \cite{HCSC_loeb_2009} predicted that during the history of galaxies like the Milky Way (MW), many mergers of small galaxies could trigger the coalescing of central BHs and kick out the remnant BH, which is about $10^3{\sim}10^5\unit{M_{\odot}}$. About dozens to hundreds of IMBHs can carry a mixture of gas and stars and form the HCSCs with a ${\lesssim}\,1\unit{pc}$ radius, fleeting in the halo of the MW-like galaxy. Rather than searching HCSCs in the halo of the MW with an all-sky survey, it may be possible to search for HCSCs in the halo of M\,31. For an HCSC with a stellar mass of $10^5\unit{M_{\odot}}$, the absolute magnitude is ${-}10^m \lesssim M_V \lesssim {-}7^m$ \citep{HCSC_merritt_2009},  which corresponds to an apparent magnitude of $15^m \lesssim m_V \lesssim 18^m$ at the distance of M\,31. The radius of this HCSC is about $0.26''$, which is resolvable with high-resolution instruments such as HST, CSST, and Euclid. It is also larger than the resolution of Gaia, therefore it might show the feature of an extended source in Gaia EDR3. Therefore, if we can eliminate contamination using radial velocities and high-resolution images, it may be possible to catch  HCSCs of M\,31 among our current candidates or new candidates selected by similar methods in the future.
\section*{acknowledgments}
The authors sincerely thank the anonymous referee for the useful comments. 
This work is supported by National Science Foundation of China (NSFC) under grant numbers 11988101/11933004/12222301, National Key Research and Development Program of China (NKRDPC) under grant numbers 2019YFA0405504/2019YFA0405500
, and Strategic Priority Program of the Chinese Academy of Sciences under grant number XDB41000000. 
This work has made use of data products from the $Gaia$, LAMOST, and PS1. 
Guoshoujing Telescope (the Large Sky Area Multi-Object Fiber Spectroscopic Telescope LAMOST) is a National Major Scientific Project built by the Chinese Academy of Sciences. 
Funding for the project has been provided by the National Development and Reform Commission. 
LAMOST is operated and managed by the National Astronomical Observatories, Chinese Academy of Sciences. We acknowledge the science research grants from the China Manned Space Project with NO. CMS-CSST-2021-A08 and CMS-CSST-2021-A09. 
\bibliography{wyl2023a}{}
\bibliographystyle{aasjournal}
\end{document}